\documentclass[11pt]{article}

\usepackage{amsmath}
\usepackage{times}
\usepackage{bm}
\usepackage{amsfonts}
\usepackage{authblk}
\usepackage{color}
\usepackage{comment}
\usepackage{natbib}
\usepackage{multirow}
\usepackage{xr}
\usepackage{array}
\usepackage{graphicx}
\usepackage{tabularx}
\usepackage{url}
\usepackage[plain,noend]{algorithm2e}

\usepackage{caption,subcaption}

\newtheorem{definition}{Definition}[section]
\newtheorem{theorem}{Theorem}[section]
\newtheorem{lemma}{Lemma}[section]

\newtheorem{remark}{Remark}[section]

\newtheorem{proposition}{Proposition}[section]
\newtheorem{algo}{Algorithm}[section]
\newtheorem{claim}{Claim}

\newcommand{\RHT}{\mathrm{RHT}}
\newcommand{\mE}{\mathbb{E}}
\newcommand{\mR}{\mathbb{R}}

\newcommand{\mF}{m_{F_{n,p}}}
\newcommand{\mP}{\mathbb{P}}
\newcommand{\cov}{\mbox{Cov}}
\newcommand{\var}{\mbox{Var}}
\newcommand{\tr}{\mbox{tr}}
\newcommand{\X}{\mathbf{X}}
\newcommand{\tS}{\tilde{S}_n}
\newcommand{\Z}{\mathbf{Z}}
\newcommand{\A}{A}
\newcommand{\ARHT}{\mathrm{ARHT}}
\newcommand{\WRHT}{\mathrm{WRHT}}
\newcommand{\B}{B}
\newcommand{\D}{D}
\DeclareMathOperator*{\argmax}{arg\,max}

\newcommand{\PreserveBackslash}[1]{\let\temp=\\#1\let\\=\temp}
\newcolumntype{C}[1]{>{\PreserveBackslash\centering}p{#1}}
\newcolumntype{L}[1]{>{\raggedright\let\newline\\\arraybackslash}p{#1}}

\begin{document}

\allowdisplaybreaks


\title{An adaptable generalization of Hotelling's $T^2$ test \\in high dimension}

\author{Haoran Li\footnote{Li's research was partially supported by NSF grant DMS 1407530.}}
\author{Alexander Aue\footnote{Aue's research 	was partially supported by NSF grants DMS 1305858 and DMS 1407530.}}
\author{Debashis Paul\footnote{Paul's research was partially supported by NSF grant DMS1407530 and NIH grant 1R01EB021707.}}
\author{Jie Peng\footnote{Peng's research was partially supported by NSF grant DMS-1148643 and NIH grant 1R01EB021707.}}
\affil{University of California, Davis}
\author{Pei Wang\footnote{Wang's research was partially supported by NIH grants R01GM108711 and U24 CA210993.}}
\affil{School of Medicine at Mount Sinai}

\date{\today}
\maketitle

%
%
%
%
%
%
%


\begin{abstract}
We propose a two-sample test for detecting the difference between mean vectors in a high-dimensional regime based on a ridge-regularized Hotelling's $T^2$. To choose the regularization parameter, a method is derived that aims at maximizing power within a class of local alternatives. We also propose a composite test that combines the optimal tests corresponding to a specific collection of local alternatives. Weak convergence of the stochastic process corresponding to the ridge-regularized Hotelling's $T^2$ is established and used to derive the cut-off values of the proposed test. Large sample properties are verified for a class of sub-Gaussian distributions. Through an extensive simulation study, the composite test is shown to compare favorably against a host of existing two-sample test procedures 
in a wide range of settings. 
The performance of the proposed test procedures is illustrated through an application to a breast cancer data set where the goal is to detect the pathways with different DNA copy number alterations across breast cancer subtypes.



\end{abstract}


\section{Introduction}
The focus of this paper is on the classical problem of testing for the equality of means of  two populations having an unknown but equal covariance matrix,  when dimension is comparable to sample size. 
The standard solution to the two-sample testing problem is the well-known Hotelling's $T^2$ test \citep{Anderson1984,Muirhead1982}. In spite of its central role in classical multivariate statistics, Hotelling's $T^2$ test has several limitations when dealing with data whose dimension $p$ is comparable to, or larger than, the sum $n=n_1+n_2$ of the two sample sizes $n_1$ and $n_2$. The test statistic is not defined for $p>n$ because of the singularity of the sample covariance matrix, but the test is also known to perform poorly in cases for which $p<n$ with $p/n$ close to unity. For example, \citet{BaiS1996} showed that the test is inconsistent in the asymptotic regime $p/n\to\gamma\in(0,1)$. 


Many approaches have been proposed in the literature to correct for the inconsistency of Hotelling's $T^2$ in high dimensions. One approach seeks to construct modified test statistics based on replacing the quadratic form involving the inverse sample covariance matrix with appropriate estimators of the squared distance between  (rescaled) population means 
\citep{BaiS1996, SrivastavaD2008, Srivastava2009, DongEtAl2016, ChenQ2010}. 
A different approach involves considering 
random projections of
the data into a certain low-dimensional space and then using  the Hotelling's $T^2$ statistics computed from the projected
data \citep{LopesJW2011,SrivastavaLR2016}.

Among other approaches to the problem under the ``dense alternative'' setting, \citet{BiswasGhosh2014} considered nonparametric, graph-based two-sample tests 
and \cite{ChakrabortyC2017} robust testing procedures. 
A different line of research involves assuming certain forms of sparsity for the difference of mean vectors.  \cite{CaiLX2014}
used this framework, in addition assuming that a ``good'' estimate of the precision matrix 
is available, and constructed tests based on the maximum component-wise mean difference of suitably transformed observations.
\citet{XuLinWeiPan2016} proposed an adaptive two-sample test based on the class of $\ell_q$-norms of the difference between sample means. 
Other recent contributions exploiting sparsity assumptions in high dimensions include \citet{WangPengLi2015}, 
\citet{GregoryCBL2015}, 
\citet{ChenLiZhong2014},
\citet{ChangZhouZhou2014}, 
and \citet{GuoChen2016}. 

In this paper, we work under the scenario $p/n \to \gamma \in (0,\infty)$, assuming that the two sample sizes are asymptotically proportional. The proposed test statistic is built upon the \textit{Regularized Hotelling's $T^2$ (RHT)} statistic introduced in \cite{ChenPPW2011} for the one-sample case, but significantly extends its scope. 
The first major contribution of this work is to provide a data-driven selection mechanism for the regularization parameter based on maximizing power under local alternatives. 
Specific focus is on a class of probabilisitic alternatives described in terms of a sequence of priors for the difference $\mu$ in the population mean vectors. 
Determination of the optimal regularization parameter does not require any knowledge of $\Sigma$.
We also show that the test of \cite{BaiS1996} is a limit of a minimax RHT test with respect to a specific class of priors.



The second main contribution is the construction of a new composite test by combining the RHT statistics corresponding to a set of optimally chosen regularization parameters. 
This data-adaptive selection of $\lambda$ allows the proposed test to have excellent power characteristics under various scenarios, such as different levels of decay of eigenvalues of $\Sigma$, and various types of structure of $\mu$.
We validate this property through extensive simulations involving a host of alternatives covering a wide range of mean and covariance structures. 
The proposed method has excellent empirical performance even when $p$ is significantly larger than $n$.  Because of these properties, and since the prefixes ``robust'' and ``adaptive'' are already part
of the statistical nomenclature tied to specific contexts, the new composite testing procedure is termed ``adaptable RHT'', 
abbreviated as ARHT. We also establish the weak convergence of a 
normalized version of the stochastic process $(\RHT(\lambda)\colon \lambda \in C)$ to a Gaussian limit, where $C\subset\mathbb{R}_+$ is a compact interval. This result facilitates computation of the cut-off values for the ARHT test.  

As a final key contribution, we establish the asymptotic behavior of the test by relaxing the assumption of Gaussianity to sub-Gaussuanity. 
Establishing this result is non-trivial due to the lack of independence between sample mean and covariance matrix in non-Gaussian settings. 
Moreover, it is shown that a simple monotone transformation of the test statistic, or a $\chi^2$ approximation, can significantly enhance the finite-sample behavior of the proposed tests.

The rest of the paper is organized as follows. Section~\ref{sec:RHT} introduces the RHT statistic
and studies a class of local alternatives.
The adaptable RHT (ARHT) test statistic is considered in Section~\ref{sec:composite_test}. 
Section~\ref{sec:calibration} discusses 
finite-sample adjustments. 
Asymptotic analysis in the non-Gaussian case is given in Section~\ref{sec:non_gaussianity}. A simulation study is reported in Section~\ref{sec:simul} and an application 
to breast cancer data is described in Section~\ref{sec:application}. 
Section~\ref{sec:discussion} has additional discussions. Proofs of the main theorems are presented in Section~\ref{sec:proofs}, and some auxiliary results are stated in the Appendix. Further technical details and additional simulation results are collected in the Supplementary Material at \url{http://anson.ucdavis.edu/~lihaoran}. R packages ARHT can be found at \url{https://github.com/HaoranLi/ARHT}. 

\section{Regularized Hotelling's $T^2$ test}
\label{sec:RHT}

\subsection{Two-sample RHT}
\label{sec:RHT:prelim}

This section introduces the two-sample regularized Ho\-telling's $T^2$ statistic.
It is first assumed that $X_{ij}{\sim} \mathcal{N}(\mu_i,\Sigma)$, $j=1,\dots, n_i$, $i=1,2$, are two independent samples with common $p\times p$ non-negative population covariance $\Sigma \equiv \Sigma_p$. 
More general sub-Gaussian observations will be treated in Section~\ref{sec:non_gaussianity}. The matrix $\Sigma$ can be estimated by its empirical counterpart, the ``pooled'' sample covariance matrix  $S_n=(n-2)^{-1}\sum_{i=1}^2\sum_{j=1}^{n_i} (X_{ij} - \bar{X}_i)(X_{ij} - \bar{X}_i)^T$, where $n=n_1+n_2$, $\bar X_i$ is the sample mean of the $i$th sample, and ${}^T$ is used to denote transposition of matrices and vectors. This framework
has been assumed in much of the work on high-dimensional mean testing problems \citep{BaiS1996,CaiLX2014}. The proposed test procedure is applicable even when the assumption of common population covariance is violated, although implications for the power characteristics of the test will be context-specific.

Due to the singularity of $S_n$ when $p>n$, it is proposed to test 
$H_0\colon \mu_1=\mu_2$ based on the family of ridge-regularized Hotelling's $T^2$ statistics
\begin{equation}\label{eq:RHT}
\RHT(\lambda) = \frac{n_1n_2}{n_1+n_2} (\bar{X}_1-\bar{X}_2)^T (S_n + \lambda I_p)^{-1} (\bar{X}_1-\bar{X}_2),
\end{equation}
indexed by a tuning parameter $\lambda> 0$ controlling the regularization strength. Observe that taking $\lambda$ to infinity leads to the procedure of \citet{BaiS1996}. 

The limiting behavior of $\RHT(\lambda)$ is tied to the spectral properties of  $\Sigma$. Let $\tau_{1,p} \geq\cdots \geq\tau_{p,p} \geq 0$ be the eigenvalues of $\Sigma$ and $H_p(\tau)=p^{-1}\sum_{\ell=1}^p\mathbf{1}_{[\tau_{\ell,p},\infty)}(\tau)$ its Empirical Spectral Distribution (ESD). 
The following assumptions are made.
\begin{itemize}
\itemsep+.8ex
\item[\textbf{C1}]
$\Sigma_p$ is non-negative definite and  $\limsup_{p}\tau_{1p}<\infty$;

\item[\textbf{C2}] 
\textit{High-dimensional setting:} 
$p,n \to \infty$ such that $n_1/n\to\kappa\in (0,1) $,  $\gamma_n= p/n \to \gamma \in (0,\infty)$ and $\sqrt{n}|p/n - \gamma| \to 0$;

\item[\textbf{C3}]
\textit{Asymptotic stability of PSD:} 
$H_p(\tau)$ 
converges as $p \to \infty$ to a probability distribution function $H(\tau)$ at every point of continuity of
$H$, and $H$ is nondegenerate at 0. 
Moreover, $\sqrt{n}\|H_p-H\|_\infty\to0$.
\end{itemize}

Since $\lambda>0$ and in view of \eqref{eq:RHT}, it suffices in 
\textbf{C1} 
to require non-negative
definiteness of $\Sigma_p$ rather than positive definiteness. The condition $\limsup_p\tau<\infty$ is necessary
to obtain eigenvalue bounds. Condition~\textbf{C2} 
ensures a well-balanced sampling design and defines the asymptotic regime
in a way that dimensionality $p$ and sample sizes $n_1$ and $n_2$ grow proportionately. Condition~\textbf{C3} 
restricts the variability allowed in $H_p$ as $p$ increases, the $\sqrt{n}$-rate of convergence being a technical requirement needed to represent the asymptotic distribution of the normalized RHT statistics in terms of functionals of the \textit{population spectral distribution (PSD)} $H$. 


Let $I_p$ be the $p\times p$ identity matrix and, for $z\in\mathbb{C}$, denote by $R_n(z)=(S_n-zI_p)^{-1}$ and $m_{F_{n,p}}(z)=p^{-1}\tr \{R_n(z)\}$ the resolvent and \textit{Stieltjes transform} of the
ESD of $S_n$ (see, for example, \cite{BaiSilverstein2010} for more
details). 
It is well-known that, $m_{F_{n,p}}(z)$ converges pointwise almost surely on $\mathbb{C}_+=\{z=u+\imath v\colon v>0\}$ to a non-random limiting distribution with Stieltjes transform $m_F(z)$ given as solution to the equation $m_F(z)=\int[\tau\{1-\gamma-\gamma zm_F(z)\}-z]^{-1}dH(\tau)$. 
This convergence holds even when $z \in \mathbb{R}_-$ 
and $m_F$ has a smooth extension to the negative reals. 
Following the same calculations as in \cite{ChenPPW2011},
under \textbf{C1}--\textbf{C3},  asymptotic mean and variance of the two-sample $\RHT(\lambda)$ under Gaussianity, are (up to multiplicative constants), given by 
\begin{align}\label{eq:Theta_1}
\Theta_1(\lambda,\gamma)
&= \frac{1- \lambda m_F(-\lambda)}{1-\gamma\{1 - \lambda m_F(-\lambda)\}}, \\[.2cm]
\label{eq:Theta_2}
\Theta_2(\lambda,\gamma)
&=\frac{1- \lambda m_F(-\lambda)}{[1-\gamma\{1 - \lambda m_F(-\lambda)\}]^3} -
\lambda\frac{\{m_F(-\lambda) - \lambda m_F'(-\lambda)\}}{[1-\gamma\{1 - \lambda m_F(-\lambda)\}]^4}.
\end{align}
Moreover, the asymptotic normality of $\RHT(\lambda)$ can be established.

These expressions are derived by making use of the following key fact: 
for every fixed $\lambda > 0$, the random matrix $R_n(-\lambda)=(S_n+ \lambda I)^{-1}$ has a \textit{deterministic equivalent} \citep{BaiSilverstein2010,LiuAuePaul15,PaulA2014} given by
\begin{equation}
\label{eq:deterministic_equivalent}
D_p(-\lambda)=
\bigg(\frac{1}{1+\gamma\Theta_1(\lambda, \gamma)}\Sigma_p + \lambda I_p\bigg)^{-1}
\end{equation}
in the sense that, for 
symmetric matrices $A$ bounded in operator norm, 
\begin{equation}
\label{eq:deterministic_equivalent_result}
\frac{1}{p}\tr\big\{R_n(-\lambda)A\big\} -
\frac{1}{p}\tr\big\{D_p(-\lambda)A\big\}
\to 0, ~~\mbox{with probability 1, as}~n \to \infty. 
\end{equation}
These results hold more generally under the sub-Gaussian model described in Section \ref{sec:non_gaussianity}.


Suppose $\Theta_j(\lambda,\gamma)$ is replaced with its empirical version $\hat\Theta_j(\lambda,\gamma_n)$ 
by  substituting $m_F(-\lambda)$ with $m_{F_{n,p}}(-\lambda)$ and $m_F'(-\lambda)$ with $m_{F_{n,p}}'(-\lambda)
=p^{-1}\tr\{R^2_n(-\lambda)\}$. Since $\hat\Theta_j(\lambda,\gamma_n)$ are $\sqrt{p}$-consistent estimators for $\Theta_j(\lambda,\gamma)$, $j=1,2$,
the RHT test rejects the null hypothesis of equal means at asymptotic level $\alpha \in (0,1)$ if
\begin{equation}\label{eq:RHT_test_statistic}
T_{n,p}(\lambda)
= \sqrt{p}\frac{\{p^{-1}\RHT(\lambda) -\hat{\Theta}_1(\lambda,\gamma_n)\}}
{\{2\hat{\Theta}_2(\lambda,\gamma_n)\}^{1/2}}
> \xi_\alpha,
\end{equation}
where $\xi_\alpha$ is the $1-\alpha$ quantile of the standard normal distribution $\mathcal{N}(0,1)$.

\subsection{Asymptotic power}\label{subsec:asymptotic_power}

This subsection deals with the behavior of 
$\RHT(\lambda)$ under local alternatives, which is critical for the determination of an optimal 
regularization parameter $\lambda$. Defining $\mu=\mu_1-\mu_2$, consider first a sequence of alternatives satisfying
\begin{equation}\label{eq:mu_stability_condition}
\sqrt{n} \mu^T D_p(-\lambda)
\mu \to q(\lambda,\gamma)
\end{equation}
as $n\to\infty$ for some $q(\lambda,\gamma) > 0$, where $D_p(-\lambda)$ is the deterministic equivalent defined in (\ref{eq:deterministic_equivalent}). The following result determines the limit of the power function
\begin{equation}\label{eq:RHT_power_function}
\beta_n(\mu,\lambda) =\mathbb{P}_\mu\{T_{n,p}(\lambda)> \xi_\alpha\}
\end{equation}
of the $\RHT(\lambda)$ test with asymptotic level $\alpha$, where $\mathbb{P}_\mu$ denotes the distribution under $\mu$.

\begin{theorem}\label{thm:local_power}
Suppose that 
\textbf{C1}--\textbf{C3}
and 
\eqref{eq:mu_stability_condition} hold. Then, for any $\lambda>0$,
\begin{equation}\label{eq:RHT_asymptotic_power}
\beta_n(\mu,\lambda) \to
\Phi\bigg(-\xi_\alpha +
\kappa(1-\kappa)\frac{q(\lambda,\gamma)}{{\{2\gamma\Theta_2(\lambda,\gamma)\}^{1/2}}}\bigg)
\qquad(n\to\infty),
\end{equation}
where $\Phi$ denotes the standard normal CDF 
and $\Theta_2(\lambda,\gamma)$ is defined in \eqref{eq:Theta_2}.
\end{theorem}

\begin{remark}
\label{rem:mu_stability}
(a)  Let $\mathbf{E}_j$ denote the eigen-projection matrix associated with the $j$th largest eigenvalue $\tau_{j,p}$ of\/ $\Sigma_p$. Suppose that there exists a sequence of functions $f_p\colon \mathbb{R}^+\cup\{0\} \to \mathbb{R}^+\cup\{0\}$ satisfying $f_p(\tau_{j,p}) = \sqrt{n}p  \| \mathbf{E}_j \mu\|^2$, $j=1,\ldots,p$, and a function $f_{\infty}$ continuous on $\mathbb{R}^+\cup\{0\}$ such that $\int |f_p(\tau) - f_\infty(\tau)| dH_p(\tau) \to 0$ as $p \to \infty$. (A sufficient condition for the latter is that  $\|f_p - f_\infty\|_\infty \to 0$ as $p \to \infty$.) Then, it follows from 
\textbf{C3} that \eqref{eq:mu_stability_condition} holds with
\begin{align}
\label{eq:q_lambda_representation}
q(\lambda,\gamma)
&= \{1+\gamma \Theta_1(\lambda,\gamma)\}
\int\frac{f_\infty(\tau)dH(\tau)}{\tau + \lambda\{1+\gamma \Theta_1(\lambda,\gamma)\}} \\
&= \int \frac{f_\infty(\tau)dH(\tau)}{\tau\{1-\gamma(1-\lambda m_F(-\lambda))\} +\lambda}. \nonumber
\end{align}
The second line in \eqref{eq:q_lambda_representation} follows from the relationship $\{1+\gamma\Theta_1(\lambda, \gamma)\}^{-1} = 1-\gamma +\lambda \gamma  m_F(-\lambda)$, for $\lambda > 0$.

(b) If\/ $\Sigma_p = I_p$, then \eqref{eq:mu_stability_condition} is satisfied if $\sqrt{n}\|\mu\|^2 \to c^2>0$. In this case, $q(\lambda,\gamma) = c^2 \Theta_1(\lambda,\gamma)$.

\end{remark}

While deterministic local alternatives like \eqref{eq:q_lambda_representation} provide useful information, 
in the following we focus on probabilistic alternatives which provide a convenient framework for incorporating structures.
Focus is on the following class of priors for $\mu$ under the alternative hypothesis.
\begin{itemize}
\item[\textbf{PA}] Assume that, under the alternative, $\mu = n^{-1/4}p^{-1/2} B \nu$ where $B$ is a $p \times p$ matrix, and $\nu$ is random vector with independent coordinates such that $\mathbb{E}[\nu_i]=0$, $\mathbb{E}[|\nu_i|^2] = 1$ and $\max_{i}\mathbb{E}[|\nu_i|^4] \leq p^{c_\nu}$ for some $c_\nu \in (0,1)$. Moreover, let $\mathbf{B} = BB^T$ with $\| \mathbf{B} \| \leq C_1 < \infty$, and, as $n,p \to \infty$,
\begin{equation}\label{eq:q_lambda_probabilistic}
p^{-1} \tr\{D_p(-\lambda)\mathbf{B}\} \to q(\lambda, \gamma),
\end{equation}
for some finite, positive constant $q(\lambda, \gamma)$.
\end{itemize}	

\begin{remark}
\label{rem:PA_variance}	
To better understand \textbf{PA}, first observe that $\mu$ has zero mean and covariance matrix $n^{-1/2} p^{-1}\mathbf{B}$. The factor $n^{-1/2}p^{-1}$ provides the scaling for the RHT test to have non-trivial local power. To check 
the meaning of (\ref{eq:q_lambda_probabilistic}), similar to the analyis in Remark \ref{rem:mu_stability}, postulate the existence of functions $\tilde f_p$ satisfying $\tilde f_p(\tau_{j,p}) = \tr\{\mathbf{E}_j \mathbf{B}\}$ and 
$\int |\tilde f_p(\tau) - f_\infty(\tau)| dH_p(\tau) \to 0$ for some function $f_\infty$ continuous on $\mathbb{R}^+\cup\{0\}$. Then, the limit in (\ref{eq:q_lambda_probabilistic}) exists and the corresponding $q(\lambda,\gamma)$ has the form given in \eqref{eq:q_lambda_representation}. Thus, $f_\infty$ can be viewed as a distribution of the total spectral mass of\/ $\mathbf{B}$ (measured as $\tr\{\mathbf{B}\}$) across the eigensubspaces of $\Sigma_p$.
\end{remark}

The framework \textbf{PA} is quite general, encompassing both dense and sparse alternatives, as illustrated in the following special cases.
\begin{itemize}
\itemsep+.6ex
\item[(I)] \textit{Dense alternative:} $\nu_i \stackrel{i.i.d.}{\sim} \mathcal{N}(0,1)$.
\item[(II)] \textit{Sparse alternative:} $\nu_i \stackrel{i.i.d.}{\sim} 
G_{\eta}$, for some $\eta \in (0,1)$, where $G_\eta$ is the discrete probability distribution which assigns mass $1-p^{-\eta}$
on 0 and mass $(1/2)p^{-\eta}$ on the points $\pm p^{\eta/2}$. 
\end{itemize}
If $B = I_p$ under (II), then $\mu$ is sparse, with the degree of sparsity determined by $\eta$.

\begin{theorem}\label{thm:local_power_probabilistic}
Suppose that 
\textbf{C1}--\textbf{C3} hold and that, under 
the alternative $H_a\colon\mu\neq 0$, 
$\mu$ has prior given by \textbf{PA}. Then, for any $\lambda>0$,
\begin{equation}\label{eq:RHT_asymptotic_power_probabilistic}
\beta_n(\mu,\lambda) \to
\Phi\bigg(-\xi_\alpha +
\kappa(1-\kappa)\frac{q(\lambda,\gamma)}{{\{2\gamma\Theta_2(\lambda,\gamma)\}^{1/2}}}\bigg)
\qquad(n\to\infty),
\end{equation}
where the convergence in \eqref{eq:RHT_asymptotic_power_probabilistic} holds in the $L^1$-sense.
\end{theorem}

\begin{remark}
\label{rem:non_convergence_mu}
Note that, even if the quantity $q_p(\lambda,\gamma) =p^{-1}\tr\{D_p(-\lambda)\mathbf{B}\}$ in \eqref{eq:q_lambda_probabilistic} does not converge, it can be verified that the difference between the left- and right-hand sides of \eqref{eq:RHT_asymptotic_power_probabilistic} still converges (in $L^1$) to zero if $q(\lambda,\gamma)$ is replaced by $q_p(\lambda,\gamma)$.
\end{remark}

Theorem \ref{thm:local_power_probabilistic} notably shows that, even for alternatives that are sparse in the sense of (II), the proposed test has the same asymptotic power as for the dense alternatives (I), as long as the covariance structure is the same. The local power of the RHT test can be compared to a test based on maximizing coordinate-wise $t$-statistics \citep[as in][]{CaiLX2014} under the sparse alternatives (II). For simplicity, let $\mathbf{B} = I_p$ and $\Sigma = I_p$.  If $\eta \in (0,1/2)$, then the size of each spike of the vector $\mu$ is of order $n^{-1/4}p^{-1/2+\eta/2} = o(n^{-1/2})$, while the maximum of the $t$-statistics is at least of the order $O_P(n^{-1/2})$ under the null hypothesis. This renders procedures based on maxima of $t$-statistics ineffective, while RHT still possesses non-trivial power. However, if $\eta > 1/2$, corresponding to a high degree of sparsity, tests based on maxima of $t$-statistics will outperform RHT. This characteristic of the RHT test is shared by the test of \cite{ChenQ2010}.

\subsection{Data-driven selection of $\lambda$}
\label{sec:lambda_selection}

Given a sequence of local probablistic alternatives, the strategy is to choose $\lambda$ by maximizing the ``local power'' function $\beta_n(\mu,\lambda)$. 
Theorems \ref{thm:local_power}  and \ref{thm:local_power_probabilistic} suggest that $\lambda$ should be chosen such that the ratio $Q(\lambda,\gamma)=q(\lambda,\gamma){\{\gamma\Theta_2(\lambda,\gamma)\}^{-1/2}}$ is maximized, with $q(\lambda, \gamma)$ given by 
(\ref{eq:q_lambda_probabilistic}). In the following, we present some
settings where $q(\lambda,\gamma)$ can be computed explicitly.
%
More specifically, two possible scenarios were considered under \textbf{PA}. (i) Suppose that $\mathbf{B}$ is specified. In this case, $q(\lambda,\gamma)$ is estimated by $p^{-1} \tr( (S_n +\lambda I_p)^{-1} \mathbf{B})$, the latter being a consistent estimator of the LHS of (\ref{eq:q_lambda_probabilistic}). (ii) Only 
the spectral mass distribution of $\mathbf{B}$ in the form of $f_\infty$ (described in Remark \ref{rem:PA_variance}) is specified. The remainder of this subsection is devoted to dealing with this scenario.


Even if \textbf{PA} holds and $f_\infty$ is specified, the computation of $q(\lambda,\gamma)$ using \eqref{eq:q_lambda_representation} remains challenging since the latter involves the unknown PSD $H$.
In order to estimate $q(\lambda, \gamma)$, without having to estimate $H$, it is convenient to have it in a closed form. This is feasible if $f_\infty$ is a polynomial. The latter is true if $\mathbf{B}$ is a matrix polynomial in $\Sigma$. Since any arbitrary smooth function can be approximated by polynomials, this formulation is quite useful and fairly
general.

Under the alternative, the following model is therefore assumed: $\mu$ satisfies \textbf{PA} with $\mathbf{B} = \sum_{m=0}^r \pi_m \Sigma^m$, for 
pre-specified 
$\pi_0,\pi_1,\ldots,\pi_r$ such that $\mathbf{B}$ is positive semidefinite. Then,
\begin{equation} \label{eq:mu_model}
\mbox{Var}(\mu) = \frac{1}{p\sqrt{n}} \sum_{m=0}^r \pi_m \Sigma^m.
\end{equation}
Thus, this model assumes that $\mbox{Var}(\mu)$ has a finite-order power expansion in $\Sigma$. 
We denote the prior $\mu \sim N(0,B)$ with $\mathbf{B}$ as in (\ref{eq:mu_model}) by $\mathcal{P}_{\tilde{\pi}}$.
Note that, in order for $\mathbf{B}$ to be positive semi-definite, it suffices that the real-valued polynomial $\sum_{m=0}^r \pi_m x^m$ is nonnegative on $[0,\|\Sigma\|]$. Unless $\Sigma=I_p$ or $\pi_0 = 1$, such a prior implies
a certain distribution of the coefficients of $\mu$ in the spectral coordinate system. 
Specifically, larger values of $\pi_m$ for higher powers $m$ imply that $\mu$ has larger contribution from the leading eigenvectors of $\Sigma$.

Under model \eqref{eq:mu_model}, \eqref{eq:q_lambda_probabilistic} is satisfied and the limit $q(\lambda,\gamma)$ equals
\begin{equation}\label{eq:q_lambda_pi_expr}
q(\lambda,\gamma) = \sum_{m=0}^r \pi_m
\rho_m(-\lambda,\gamma),
\end{equation} 
with $\rho_m(-\lambda,\gamma)$ satisfying the recursive formula
\[
\rho_{m+1}(-\lambda,\gamma)
= \{1+\gamma\Theta_1(\lambda,\gamma)\}
\bigg\{\int\tau^mdH(\tau) -\lambda\rho_m(-\lambda,\gamma)\bigg\}.
\]
This formula, which can be deduced from Lemma 3 of \cite{LedoitP2011}, and the derivations in the Supplementary Material, involves the population spectral moments $\int\tau^mdH(\tau)$. The latter can be estimated, since equations connecting the moments of $H$ with the limits of the tracial moments $p^{-1}\tr\{S_n^m\}$, $m\geq 1$, are known \citep[see Lemma \ref{lem:spectral_moment}, quoted from][]{BaiCY2010}.


In practice, we restrict to the case $r=2$. There are several considerations that guided this choice of $r$.  First, for $r=2$, all quantities involved can be computed explicitly without requiring knowledge of higher order moments of the observations. Also, the corresponding estimating equations for $q(\lambda, \gamma)$ are more stable as they do not involve higher order spectral moments. Secondly, the choice of $r=2$ yields a significant, yet nontrivial, concentration of the prior covariance of $\mu$ (equivalently, $\mathbf{B}$) in the directions of the leading eigenvectors of $\Sigma$. Finally, the choice $r=2$ allows for both convex and concave shapes for the spectral mass distribution
$f_\infty$ since the latter becomes a quadratic function.

With $r=2$, in order to estimate $q(\lambda, \gamma)$, it suffices to estimate
\begin{eqnarray}\label{eq:rho_0to2}
\rho_0(-\lambda,\gamma) &=& ~m_F(-\lambda),\\
\rho_1(-\lambda,\gamma) &=& ~\Theta_1(\lambda,\gamma), \nonumber\\
\rho_2(-\lambda,\gamma) &=& ~\{1+\gamma\Theta_1(\lambda,\gamma)\}\{\phi_1 - \lambda\rho_1(-\lambda,\gamma)\}, \nonumber
\end{eqnarray}
where $\phi_1=\int\tau dH(\tau)$.
The latter can be estimated accurately by $\hat\phi_1=p^{-1}\tr\{S_n\}$
(see Proposition \ref{prop:converg_phi}).
%
%
%
%
In the following the algorithm for the data-driven selection of the regularization parameter $\lambda$ is stated.

\begin{algo}[Empirical selection of $\lambda$] Perform the following steps.
\label{algorithm}
\vspace{-.5ex}
\begin{enumerate}
\itemsep-0ex
\item Specify prior weights $\tilde{\pi}=(\pi_0,\pi_1,\pi_2)$;
\item For each $\lambda$, compute the estimates 
\begin{align*}
\hat\rho_0(-\lambda,\gamma_n)&=\mF(-\lambda), \\
\hat\rho_1(-\lambda,\gamma_n)&=\hat{\Theta}_1(\lambda,\gamma_n),\\
\hat\rho_2(-\lambda,\gamma_n)&=\{1+\gamma_n\hat\Theta_1(\lambda,\gamma_n)\}\{\hat\phi_1 -\lambda\hat{\rho}_1(-\lambda,\gamma_n)\};
\end{align*}
\item For each $\lambda$, compute the estimate 
\[
\hat Q_n(\lambda,\gamma_n;\tilde{\pi})=\sum_{m=0}^2\pi_m\hat\rho_m(-\lambda,\gamma_n)/\{\gamma_n\hat\Theta_2(\lambda,\gamma_n)\}^{1/2};
\]
\item Select 
$\lambda_{\tilde\pi} \equiv \lambda_{\tilde\pi,n}=\argmax_{\lambda}\hat Q_n(\lambda,\gamma_n;\tilde\pi)$ through a grid search. 
\end{enumerate}
\end{algo}

Although in theory arbitrarily small positive $\lambda$ are allowed in the test procedure, in practice, meaningful lower and upper bounds 
$\underline{\lambda}$ and $\overline{\lambda}$ 
are needed to ensure stability of the test statistic when $p \approx n$ or $p > n$.
The recommended choices are $\underline{\lambda} = p^{-1}\tr\{S_n\}/100$ and $\overline{\lambda}=20\|S_n\|$. 

The behavior of the test with the data-driven tuning parameter is described in the following theorem.

\begin{theorem}\label{thm:converg_lambda}
Let $[\underline{\lambda},\overline{\lambda}]$ (with $\overline{\lambda} > \underline{\lambda}>0$) be a non-empty interval. Let $\lambda_\infty$ be any local maximizer of $Q(\lambda,\gamma;\tilde{\pi})$ on $[\underline{\lambda},\overline{\lambda}]$. If conditions~\textbf{C1}--\textbf{C3}
are satisfied and if there is a $C>0$ such that $\partial^2Q(\lambda_\infty,\gamma;\tilde{\pi})/\partial\lambda^2<-C$, then there exists a sequence $(\lambda_n\colon n\in\mathbb{N})$ of local maximizers of $(\hat{Q}_n(\lambda,\gamma_n;\tilde{\pi})\colon n\in\mathbb{N})$, satisfying
\begin{equation}
\label{eq:converg_lambda}
n^{1/4}|\lambda_n-\lambda_\infty|=O_p(1)
\qquad(n\to\infty).
\end{equation}
Further, under the null hypothesis,
\begin{equation}
\label{eq:normality_lambda_n}
T_{n,p}(\lambda_n)
= \frac{{p^{1/2}}\{p^{-1}\RHT(\lambda_n) -\hat{\Theta}_1(\lambda_n,\gamma_n)\}}
{\{2\hat{\Theta}_2(\lambda_n,\gamma_n)\}^{1/2}}\Longrightarrow \mathcal{N}(0,1)
\qquad(n\to\infty),
\end{equation}
where $\Longrightarrow$ denotes convergence in distribution. 
The procedure is adaptive in the sense that the asymptotic power of the test based on $T_{n,p}(\lambda_n)$ is the same as that of $T_{n,p}(\lambda_\infty)$ under the sequence of priors specified by $\tilde{\pi}$.
\end{theorem}
\begin{remark}
In Theorem \ref{thm:converg_lambda}, if $\lambda_\infty$ is a boundary point and ${\partial}Q(\lambda_\infty,\gamma;\tilde{\pi})/{\partial\lambda}\neq0$, then the assumption on ${\partial^2}Q(\lambda_\infty,\gamma;\tilde{\pi})/{\partial\lambda^2}$ can be dropped. 	
\end{remark}

\subsection{Minimax selection of $\lambda$}\label{subsec:minimax_RHT}

In Section \ref{sec:lambda_selection}, it is assumed that a specific prior $\tilde\pi$ is available. However, in practice, rather than a particular choice of $\tilde\pi$, we may have to consider a collection
of such priors. In this subsection, a procedure for selecting the regularization parameter for the RHT test $T_{n,p}(\lambda)$ is presented 
that is based on the principle of minimaxity. Throughout this subsection, \textit{minimax} refers
to minimaxity within the class of all RHT tests.
 
Let $\mathcal{D} = \{T_{n,p}(\lambda)\colon \lambda \in [\underline{\lambda},\overline{\lambda}]\}$, for $0 < \underline{\lambda} < \overline{\lambda} <\infty$ denote a class of normalized RHT test statistics. Also, let $\mathfrak{P}$ be a family of local priors for $\mu$ under the alternative. Notice that, for any $\alpha \in (0,1)$ the test 
$\delta_{\alpha}(\lambda) = \mathbf{1}(T_{n,p}(\lambda) > \xi_\alpha)$ has asymptotically level $\alpha$. For any given prior $\mathcal{P}$ for $\mu$ under the alternative, define the \textit{asymptotic Bayes risk} of the test $\delta_{\alpha}(\lambda)$ with respect to prior $\mathcal{P}$ as 
\begin{equation}\label{eq:Bayes_risk}
R(\delta_{\alpha}(\lambda); \mathcal{P}) = \limsup_{n,p\to \infty} (1-\mathbb{E}_{\mathcal{P}}[\beta_n(\mu,\lambda)]) = 1 - \liminf_{n,p\to \infty}
\mathbb{E}_{\mathcal{P}}[\beta_n(\mu,\lambda)]
\end{equation}
with $\beta_n(\mu,\lambda)$ as in (\ref{eq:RHT_power_function}).
We say that $T_{n,p}(\lambda_*)$ is a \textit{locally asymptotically minimax (LAM)} test within the class 
$\mathcal{D}$ and with respect to 
$\mathfrak{P}$, if for each $\alpha \in (0,1)$, the minimum value
of $\sup_{\mathcal{P}\in \mathfrak{P}} R(\delta_{\alpha}(\lambda); \mathcal{P})$ over $\lambda \in [\underline{\lambda},\overline{\lambda}]$ is attained at $\lambda_*$.

In the following, consider a family of priors $\mathfrak{P}_r(C)$ defined in the following way. For a constant $C > 0$, define 
\[
\Pi_{r}(C) = \{\tilde\pi = (\pi_0,\ldots,\pi_r)\colon
\sum_{m=0}^r \pi_m x^m \geq 0~\mbox{for}~ x\in [0,\infty), \;
\sum_{m=0}^r \pi_m \phi_m = C\},
\]
where $\phi_m = \int \tau^m dH(\tau)$. 
Let $\mathcal{P}_{\tilde{\pi}}$ denote the prior for $\mu$ satisfying \textbf{PA} and (\ref{eq:mu_model}).  
Finally, let
\[
\mathfrak{P}_r(C) = \{\mathcal{P}_{\tilde \pi}\colon \tilde \pi \in 
\Pi_{r}(C) \}.
\]
The condition $\sum_{m=0}^r \pi_m x^m \geq 0$ for all $x \geq 0$ ensures that the matrix $\sum_{m=0}^r \pi_m \Sigma^m$ is non-negative definite, while the condition
$\sum_{m=0}^r \pi_m \phi_m = C$ means that as $p\to \infty$, $\sqrt{n}\tr\{\mbox{Var}(\mu)\} \to C$.
Observe that, for $\tilde\pi \in \Pi_{r}(C)$, the asymptotic Bayes risk $R(\delta_{\alpha}(\lambda); \mathcal{P}_{\tilde{\pi}})$ equals 
$1- \Phi(-\xi_\alpha + \kappa(1-\kappa) Q(\lambda,\gamma;\tilde{\pi}))$
where $q(\lambda,\gamma) \equiv q(\lambda,\gamma;\tilde{\pi})$ is given 
by (\ref{eq:q_lambda_pi_expr}), implying that $\mathcal{P}_{\tilde{\pi}}$ actually constitutes an equivalence class of priors.

In the following restricting to $r=2$, note that finding 
an LAM test within the class $\mathcal{D}$ and with respect to the family $\mathfrak{P}_{2}(C)$
means finding a $\lambda \in [\underline{\lambda},\overline{\lambda}]$ that minimizes $\sup_{\tilde{\pi}\in \Pi_2(C)} R(\delta(\lambda); \mathcal{P}_{\tilde{\pi}})$.  Without loss of generality, take $C=1$ since the risk function is monotonically decreasing in $Q(\lambda,\gamma;\tilde{\pi})$, and the latter is a linear function of $\tilde{\pi}$. This leads to the following result.  
\begin{proposition}\label{prop:RHT_LAM}
Under the conditions of Theorem \ref{thm:local_power_probabilistic}, the LAM test within the class $\mathcal{D}$, with respect to the family $\mathfrak{P}_{2}(C)$ is $T_{n,p}(\bar\lambda)$.
\end{proposition}
Proof of this proposition is given in Section \ref{subsec:proof_RHT_LAM}.

It can be verified that as $\lambda \to \infty$, the test statistic RHT$(\lambda)$  converges pointwise to the corresponding test statistic by \cite{BaiS1996},
and the local asymptotic power of RHT$(\lambda)$ 
under the class of alternatives $\mathfrak{P}_{2}(C)$ also converges to the corresponding power for the test by \cite{BaiS1996}. Thus, Proposition 
\ref{prop:RHT_LAM} shows that the test by \cite{BaiS1996} is the limit of a locally asymptotically minimax test, namely the test $T_{n,p}(\bar\lambda)$, as $\bar{\lambda} \to \infty$.

\section{Adaptable RHT}\label{sec:composite_test}

Section~\ref{sec:lambda_selection} describes a data-driven procedure for selecting the optimal regularization parameter
$\lambda$ for pre-specified prior weights $\tilde{\pi}$, where\-as 
Section \ref{subsec:minimax_RHT} derives an asymptotically minimax RHT test with respect to a class of priors. An extensive simulation analysis 
reveals that there is a considerable variation in the shape of the power function and the value of the corresponding Bayes rule,
especially when the condition number of $\Sigma$ is relatively large.

As an alternative to the minimax approach, which 
can be overly pessimistic, instead of considering a broad collection of priors, one might consider a convenient collection of priors that are representative of certain structural scenarios. Thus adopting a mildly conservative approach, 
define a new test statistic as the maximum of the RHT statistics corresponding to a set of regularization parameters that are optimal 
with respect to a specific collection of priors. Specifically, 
we propose the following test statistic, referred to as \textit{Adaptable RHT (ARHT)}:
\begin{equation}\label{eq:RHT_composite_test}
\ARHT_{n,p}(\Pi) = \max_{\tilde{\pi} \in \Pi} T_{n,p}(\lambda_{\tilde{\pi}}),
\end{equation}
where $T_{n,p}(\lambda)$ is 
defined in \eqref{eq:RHT_test_statistic}, $\lambda_{\tilde{\pi}}$ 
in Algorithm~\ref{algorithm}, and $\Pi=\{\tilde{\pi}_1,\cdots, \tilde{\pi}_k\}$, $k\geq 1$,  is a pre-specified finite class of weights. A simple but effective choice of $\Pi$ consists of the three \textit{canonical weights} $\tilde{\pi}=(1,0,0)$, $(0,1,0)$ and $(0,0,1)$. We focus on this particular specification of $\Pi$, since a convex combination of these three weights cover a wide range of local alternatives, and this choice leads to very satisfactory empirical performance as is illustrated through simulations in Section \ref{sec:simul}. In particular, the ARHT procedure is shown to 
outperform the test by \cite{BaiS1996} (the limiting LAM procedure) in most circumstances.

Determining the cut-off values of $\ARHT_{n,p}(\Pi)$ requires knowing the asymptotic distribution of the process $T_{n,p}=(T_{n,p}(\lambda)\colon\lambda\in[\underline{\lambda},\overline{\lambda}])$ under the null hypothesis of equal means. From this, the case where $\Lambda=\{\lambda_{\tilde{\pi}_1},\ldots,\lambda_{\tilde{\pi}_k}\}$ is a collection of finitely many regularization parameters can be easily derived. 
\begin{theorem}
\label{thm:convergence_process}
If 
\textbf{C1}--\textbf{C3}
are satisfied, then, under $H_0$,
\[
T_{n,p}\stackrel{d}{\longrightarrow} Z
\qquad (n\to\infty),
\]
where $\stackrel{d}{\longrightarrow}$ denotes weak convergence in the Skorohod space $D[\underline{\lambda},\overline{\lambda}]$ and $Z=(Z(\lambda)\colon \lambda\in[\underline{\lambda},\overline{\lambda}])$ 
a centered Gaussian process with covariance function
\begin{equation}\label{eq:Gamma_lambda_12_expr}
\Gamma(\lambda,\lambda^\prime) =
\{1+\gamma\Theta_1(\lambda,\gamma)\}\{1+\gamma\Theta_1(\lambda^\prime,\gamma)\}\frac{\lambda^\prime\Theta_1(\lambda^\prime,\gamma) -
\lambda\Theta_1(\lambda,\gamma)}{(\lambda^\prime - \lambda)
\{\Theta_2(\lambda,\gamma)\Theta_2(\lambda^\prime,\gamma)\}^{1/2}}, 
\end{equation}
for $\lambda\neq\lambda^\prime$, and $\Gamma(\lambda,\lambda)\equiv 1$.
In particular, for every $k\geq1$ and every collection $\Lambda=\{\lambda_1,\dots,\lambda_k\}\subset[\underline{\lambda},\overline{\lambda}]$, it holds that
\begin{equation*}
(T_{n,p}(\lambda_1),\ldots,T_{n,p}(\lambda_k))^T
\Longrightarrow
N_k(0,\Gamma(\Lambda))
\qquad (n\to\infty),
\end{equation*}
where the limit on the right-hand side is a $k$-dimensional centered normal distribution with $k\times k$ covariance matrix $\Gamma(\Lambda)$ with entries 
$\Gamma(\lambda_i,\lambda_j)$, $i,j=1,\ldots,k$.
\end{theorem}

Theorem \ref{thm:convergence_process} shows that 
$\ARHT_{n,p}(\Pi)$ has a non-degenerate limiting distribution under $H_0$. Theorem \ref{thm:convergence_process} can be used to determine the cut-off values of the test by deriving analytical formulae for the quantiles of the limiting distribution. Aiming to avoid complex calculations, 
a \textit{parametric bootstrap} procedure is applied to approximate the cut-off values. Specifically, $\Gamma(\Lambda)$ is first estimated by $\hat\Gamma_n(\Lambda)$, and then bootstrap replicates are generated by simulating from $N_{k}(0,\hat\Gamma(\Lambda))$, thereby leading to an approximation of the null distribution of $\ARHT_{n,p}(\Pi)$. 
A natural candidate for the covariance estimator is
\begin{align}
\label{eq:Gamma_hat_lambda_12}
&\hat\Gamma_n(\lambda,\lambda^\prime) \\
&= \{1+\gamma_n\hat\Theta_1(\lambda,\gamma_n)\}\{1+\gamma_n\hat\Theta_1(\lambda^\prime,\gamma_n)\}
\frac{\lambda^\prime\hat\Theta_{1}(\lambda^\prime,\gamma_n) -
  \lambda\hat\Theta_{1}(\lambda,\gamma_n)}{(\lambda^\prime - \lambda)
  \{\hat\Theta_{2}(\lambda,\gamma_n)\hat\Theta_{2}(\lambda^\prime,\gamma_n)\}^{1/2}}, \nonumber
\end{align}
for $\lambda\neq\lambda^\prime$ and $\hat\Gamma_n(\lambda,\lambda) \equiv 1$.
\begin{remark}
\label{remark_Gammahat}
{\rm
It should be noticed that $\hat\Gamma_n(\Lambda)$ defined through \eqref{eq:Gamma_hat_lambda_12} may not be non-negative definite even though it is symmetric. If such a case occurs, the 
resulting estimator can be projected to its closest non-negative definite matrix simply by setting the negative eigenvalues to zero. This covariance matrix estimator is denoted by $\hat \Gamma_n^+(\Lambda)$ and is used for generating the bootstraps samples. 
}
\end{remark}

\section{Calibration of Type I error probability}\label{sec:calibration}


Simulation studies reveal that the 
size of 
$\RHT$ 
tends to be slightly inflated. This is because a normal approximation is 
used to describe 
a quadratic form statistic, 
leading to skewed distributions in finite samples. Two remedies are proposed. The first
is based on a power transformation of $\RHT$, reducing skewness by calibrating higher-order terms in the test statistics. The second on choosing cut-off values of $\RHT$ based on quantiles of a normalized $\chi^2$ distribution whose first two moments match those of $\RHT$.

\subsection{Cube-root transformation}
\label{sec:calibration_cube}

In principle, any power transformation may be considered, but empirically, a near-symmetry of the null distribution is obtained by a cube-root transformation of the RHT statistic. 
Therefore restricting to this case only, an application of the $\delta$-method yields
\begin{equation}
\label{eq:cubicroot}
 \tilde{T}_{1/3}(\lambda)=
 \frac{{p^{1/2}} [\{p^{-1}\RHT(\lambda)\}^{1/3}- \hat\Theta_1^{1/3}(\lambda,\gamma_n)]}
 {({{2^{1/2}}}/{3})\hat\Theta_2^{1/2}(\lambda,\gamma_n)/\hat\Theta_1^{2/3}(\lambda,\gamma_n)}\Longrightarrow N(0,1).
\end{equation}
This gives rise to the cube-root transformed $\ARHT$ test statistic 
\[
\ARHT_{1/3}(\Pi)=\max\limits_{\tilde\pi \in\Pi}\tilde{T}_{1/3}(\lambda_{\tilde\pi}).
\]
A test based on $\ARHT_{1/3}(\Pi)$ for a finite set $\Pi$ of weight vectors can be performed by making use of the covariance kernel $\Gamma$
given in \eqref{eq:Gamma_lambda_12_expr}. $\ARHT_{1/3}$ is recommended for most practical applications since it nearly symmetrizes the null distribution of the test statistic even for moderate sample sizes. Algorithm \ref{algorithm:overall} details the composite test procedure with the recommended $\ARHT_{1/3}$ statistic. 
\begin{algo}[Cube-root transformed ARHT]
\label{algorithm:overall}
{~} \vspace{-.1cm}
\begin{enumerate}
\item Diagonalization: Compute the spectral decomposition of $S_n=P_n\Delta_nP_n^T$, apply the transformation  $\bar{Y}_1=P_n^T\bar{X}_1$, $\bar{Y}_1=P_n^T\bar{X}_1$; and run the rest with $\bar{X}_1, \bar{X}_2, S_n$ replaced by $\bar{Y}_1, \bar{Y}_2$ and $\Delta_n$;
\item For each $\tilde{\pi}$ in $\Pi$, run Algorithm \ref{algorithm} and obtain $\Lambda=\{\lambda_{\tilde{\pi}}\colon\tilde{\pi}\in\Pi\}$;
\item Compute $\hat\Gamma_n^+(\Lambda)$;
\item Generate $\varepsilon_1,\ldots,\varepsilon_B$ with $\varepsilon_b=\max_{1\leq i \leq k} Z_i^{(b)}$ with $Z^{(b)}\sim \mathcal{N}(0,\hat\Gamma_n^+(\Lambda))$;
\item Compute $\ARHT_{1/3}(\Pi)$;
\item Compute $p$-value as $B^{-1}\sum_{b=1}^B I\{\varepsilon_b>\ARHT_{1/3}(\Pi)\}$.
\end{enumerate}
\end{algo}

\subsection{$\chi^2$-approximation of cut-off values} \label{sec:chi_approximation}

While the cube-root transformation is shown to be quite effective, a weighted chi-square approximation 
can also be used to calibrate the size of $\ARHT$. This involves setting the cut-off values as quantiles of the maximum of a set of scaled $\chi^2$ distributions, i.e., random variables of the form $a\chi^2(\ell)$, where $a$ is a normalizing constant and $\ell$ is the degree of freedom.  For each pair $(a,\ell)$, the $a\chi^2(\ell)$ distribution is used to mimic the distribution of $\RHT$ in (\ref{eq:RHT}) for a given regularization parameter $\lambda$. The scale multipliers $a$ and the degrees of freedom $\ell$ are selected so that the first two moments and the covariances of the $\chi^2$ variables match with those of the corresponding $\RHT$ test.
Details are given in the Supplementary Material. Unlike the cube-root transform of Section~\ref{sec:calibration_cube}, this method only modifies cut-off values. Based on our simulations, both methods perform similar in terms of power curves. 

\section{Extension to sub-Gaussian distributions}
\label{sec:non_gaussianity}

The results presented thus far are now extended to a general class of sub-Gaussian distributions
\citep[see][]{Chatterjee2009}. 
The extension is achieved for the independent samples model
\begin{equation}
\label{non_gaussian_setting}
X_{ij}=\mu_i+\Sigma^{1/2}_pZ_{ij},
\qquad j=1,\dots,n_i,\;\; i=1,2,
\end{equation}
where $Z_{ij}=(z_{ij1},\ldots,z_{ijp})^T$ are $p$-dimen\-sio\-nal independent random vectors with i.i.d.\ entries satisfying $\mE[z_{ijk}]=0$, $\mE[z^2_{ijk}]=1$ and $\mE[ z^3_{ijk}]=0$.
To specify the distribution of $z_{ijk}$, 
introduce the following class of probability measures. 
\begin{definition}\label{def:non_gaussian}
For each $c_1, c_2>0$, let $\mathcal{L}(c_1,c_2)$ be the class of probability measures on the real line $\mathbb{R}$ that arises as laws of random variables $u(Z)$, where $Z$ is a standard normal random variable and $u$ is a twice continuously differentiable function such that, for all $x\in\mathbb{R}$,
\begin{equation}
\label{non_gaussian}
|u^\prime(x)|\leq c_1
\qquad\text{and}\qquad
 |u^{\prime\prime}(x)|\leq c_2.
\end{equation}
\end{definition}
Note that random variables in $\mathcal{L}(c_1,c_2)$ are sub-Gaussian and have continuous distribution, since $u$ is a Lipschitz function with bounded Lipschitz constant. 
The first condition in \eqref{non_gaussian} is used to control the magnitude of the variance of $u(Z)$, while the second condition is primarily for controlling the tail behavior of the statistic. 
This approach is particularly attractive as it only requires establishing appropriate upper bounds for the operator norms of the gradient and Hessian matrices of the statistic (with respect to the variables), and matching the first two asymptotic moments. However, the calculations in our setting are non-trivial since they require a detailed analysis of the resolvent of the sample covariance matrix. 

\begin{theorem}
\label{general}
All previously stated results hold if the observations $X_{ij}$ are as in \eqref{non_gaussian_setting} with the $z_{ijk}$ satisfying Definition \ref{def:non_gaussian} together with $\mE[ z_{ijk}]=0$, $\mE [z^2_{ijk}]=1$, $\mE [z^3_{ijk}]=0$, and $\Sigma_p$ satisfying conditions~\textbf{C1}--\textbf{C3}.
\end{theorem}
Key to
the proof of Theorem \ref{general} is the consideration of a modified version of $\RHT$, replacing $S_n$ with the non-centered matrix $\widetilde S_n = n^{-1} \sum_{i=1}^2\sum_{j=1}^{n_i} X_{ij}X_{ij}^T$.
Defining $U_{kl}(\lambda) = \bar X_k^T(\widetilde S_n + \lambda I_p)^{-1} \bar X_l$, $k,l=1,2$, the joint asymptotic normality of $(U_{11}(\lambda),U_{12}(\lambda), U_{22}(\lambda))$ can first be established. 
Then, a suitable transformation of variables and an appropriate use of the $\delta$-method prove the asymptotic normality of $\RHT(\lambda)$. The proof details for Theorem \ref{general} are provided in the Supplementary Material.
The derivation of the power function of the RHT test under local alternatives follows analogously.

Theorem \ref{general} is expected to hold under even more general conditions than stated above. Indeed, in the one-sample testing problem, making use of the analytical framework adopted by \cite{PanZhou2011}, asymptotic normality of  $\RHT$ can be proved when 
Definition \ref{def:non_gaussian} is replaced by a bounded fourth moment assumption that is standard in spectral analysis of large covariance matrices. However, this derivation is rather technical and not readily extended to the two-sample setting due to certain structural differences between one- and two-sample settings under non-Gaussianity.
Whether such generalizations are feasible in the present context is a topic for future research.

\section{Simulations}
\label{sec:simul}
\subsection{Competing methods}

In this section, the proposed ARHT is compared by means of a simulation study to a host of popular competing methods, including the tests introduced by \citet{BaiS1996} (BS), \citet{ChenQ2010} (CQ), \citet{LopesJW2011} (RP), and \citet{CaiLX2014} ($\mathrm{CLX.}\bm\Omega^{1/2}$ and $\mathrm{CLX.}\bm\Omega$, corresponding to the two different transformation matrices $\Omega^{1/2}$ and $\Omega=\Sigma^{-1}$). In the following, $\ARHT$, $\ARHT_{1/3}$ and $\ARHT_{\chi^2}$ denote the original, cubic-root transformed and $\chi^2$-approximated $\ARHT$ procedure introduced in Sections~\ref{sec:composite_test}, \ref{sec:calibration_cube} and \ref{sec:chi_approximation}, respectively.
\begin{figure}[htbp]
\captionsetup[subfigure]{justification=centering}
\begin{subfigure}[b]{1.2in}
\centering
\includegraphics[width=1.2in,height=1.2in]{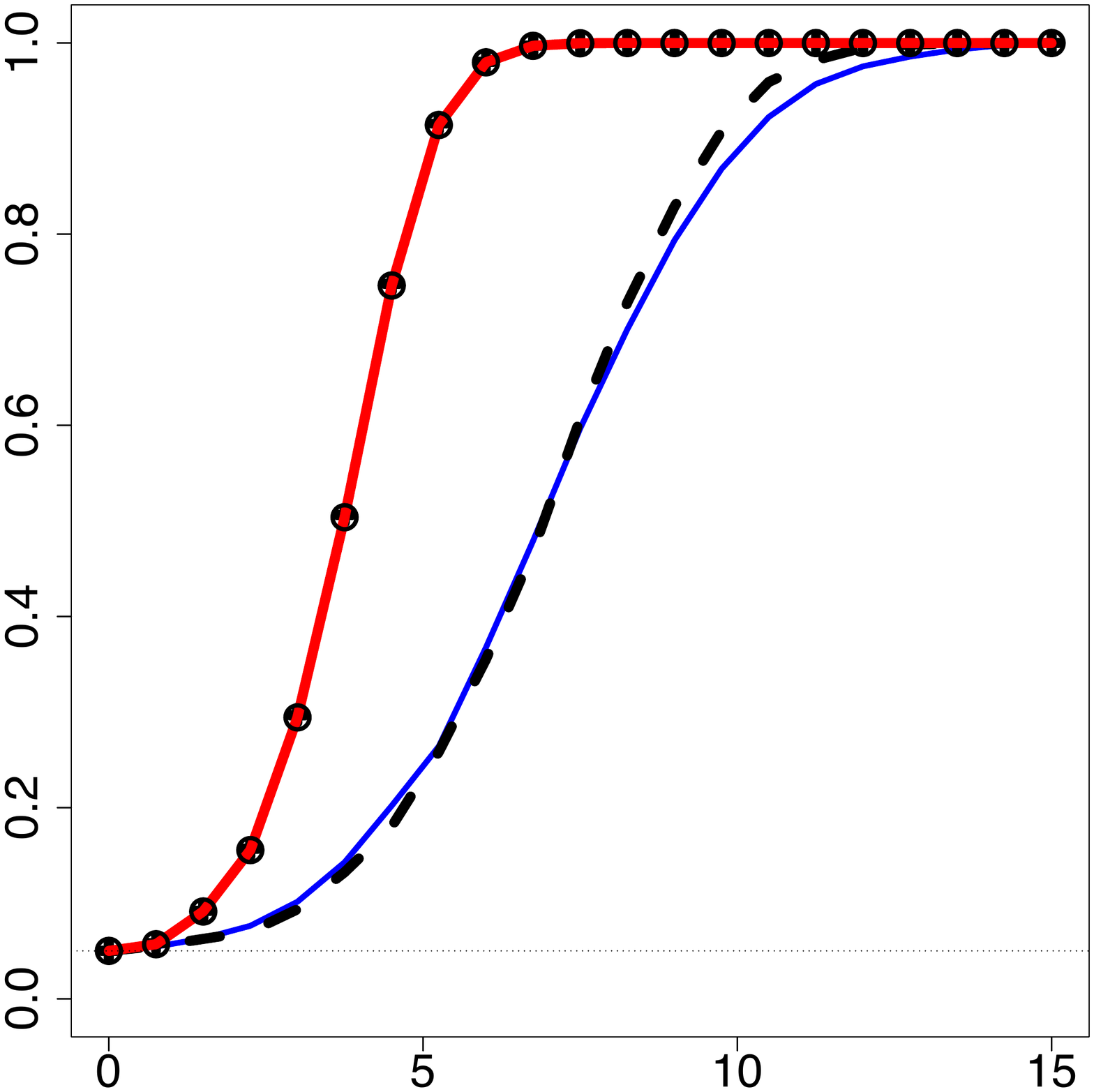}
\subcaption{p=200, \\$\mu\sim N(0,cI)$}
\end{subfigure}
\hfill
\begin{subfigure}[b]{1.2in}
\centering
\includegraphics[width=1.2in,height=1.2in]{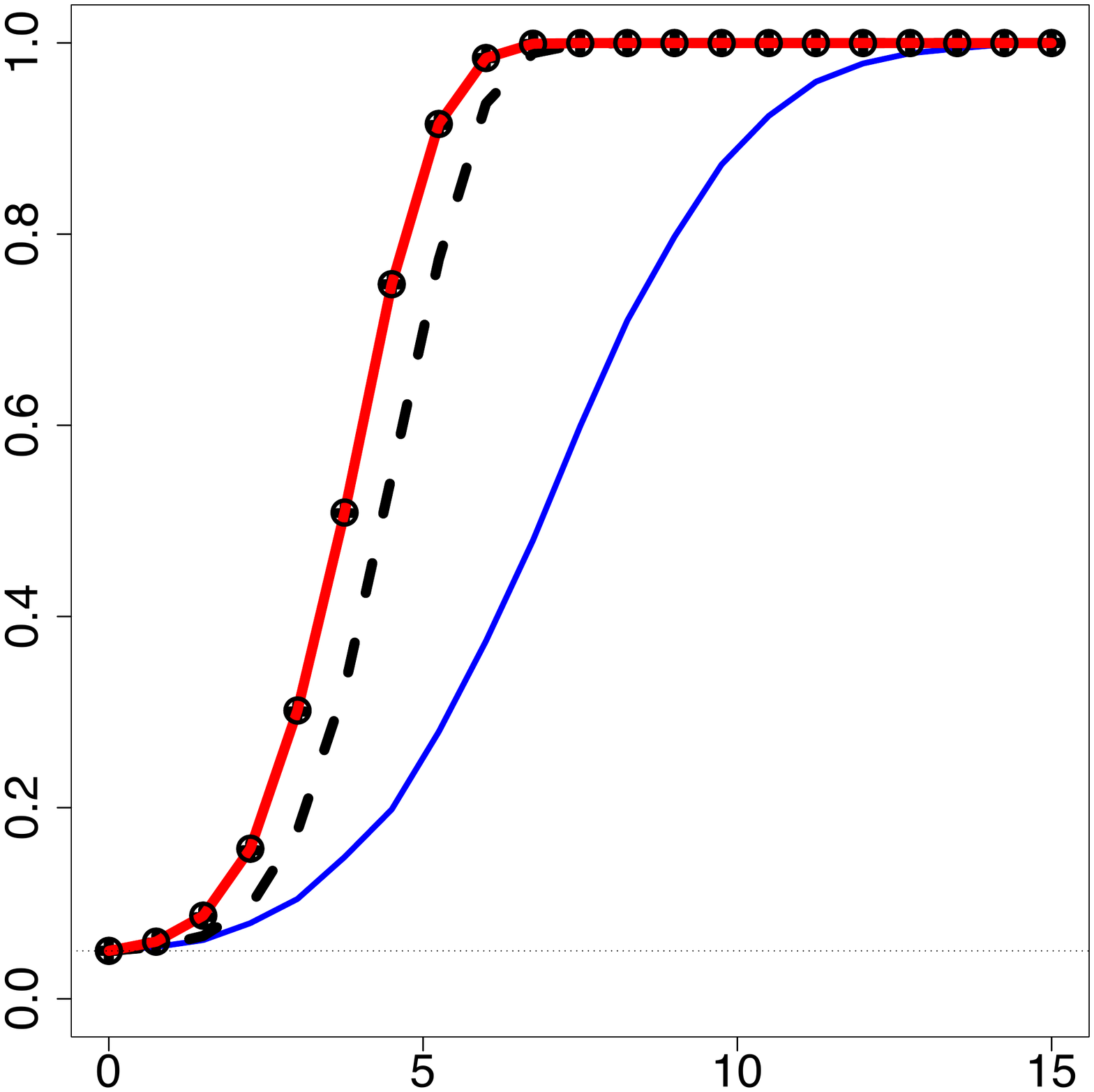}
\subcaption{p=200,\\ sparse $\mu$}
\end{subfigure}
\begin{subfigure}[b]{1.2in}
\centering
\includegraphics[width=1.2in,height=1.2in]{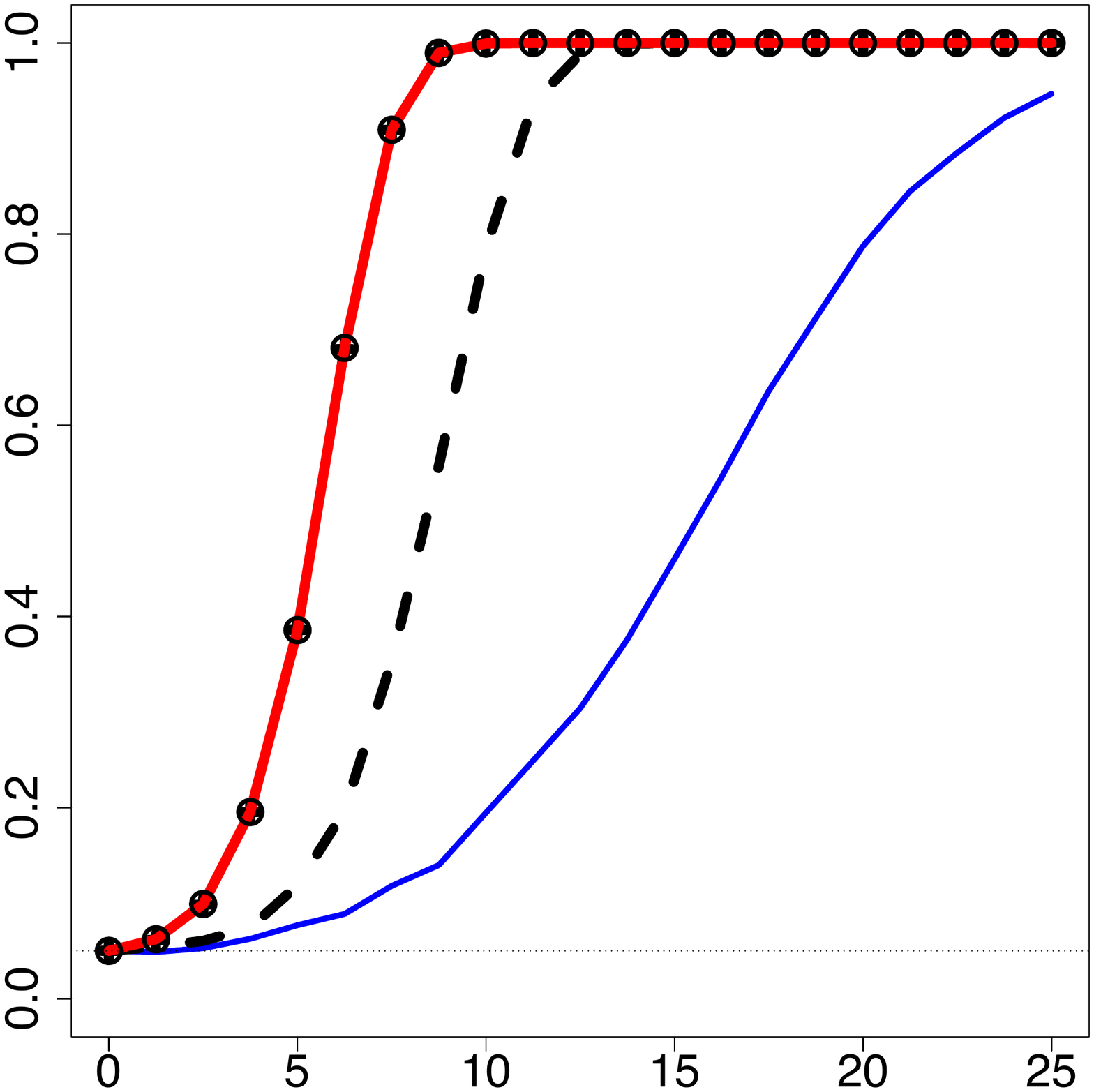}
\subcaption{p=1000,\\ $\mu\sim N(0,cI)$}
\end{subfigure}
\hfill
\begin{subfigure}[b]{1.2in}
\centering
\includegraphics[width=1.2in,height=1.2in]{AOS_size_adjusted_normal_Sigma_ID_p_1000_sparse.pdf}
\subcaption{p=1000,\\ sparse $\mu$}
\end{subfigure}
\caption{Size-adjusted empirical power with  $X_{ij}\sim N(\cdot,\Sigma)$ and  $\Sigma= \mbox{ID}$. 
$\ARHT_{1/3}$ (solid, red), $\chi^2$ approximation (circle), \citet{BaiS1996} (solid, green) \citet{ChenQ2010} (+), \citet{LopesJW2011} (solid, purple) and \citet{CaiLX2014} with $\Omega$ transform (dashed).}
\label{fig:SizeAdjusted_normal_Sigma_ID}
\end{figure}

\subsection{Settings and results}

In the simulations, the observations $X_{ij}$ are as in \eqref{non_gaussian_setting}, while two different distributions for $z_{ijk}$ are considered, namely the $N(0,1)$ distribution and the $t$-distribution with four degrees of freedom, $t_{(4)}$, rescaled to unit variance. For the normal case, the sample sizes are chosen as $n_1=n_2=50$. For the $t_{(4)}$ case, the sample sizes are chosen to be $n_1=30$ and $n_2=70$. The dimension $p$ is 50, 200, or 1000, so that $\gamma=p/(n_1+n_2)=0.5$, $2$ or 10. Results are here reported mainly for $p=200$ and 1000, while the case $p=50$ is reported in the Supplementary Material.  The range of regularization parameters is chosen as $[\underline{\lambda},\overline{\lambda}] = [0.01,100]$, using a grid with progressively coarser spacings for determining the optimal $\lambda_n \equiv \lambda_{\pi,n}$.

The following three models for the covariance matrix $\Sigma=\Sigma_p$ are considered.

(\romannumeral 1) The {\it identity matrix}\/ (ID): Here $\Sigma=I_p$;

(\romannumeral 2) The {\it sparse case}\/ $\Sigma_s$: Here $\Sigma=(p^{-1}\tr\{D\})^{-1}D$
with a diagonal matrix $D$ whose eigenvalues are given by $\tau_j=0.01+(0.1+j)^6$, $j=1,\ldots,p$;

(\romannumeral 3) The {\it dense case}\/ $\Sigma_d$: Here $\Sigma=P^T\Sigma_sP$ with a
unitary matrix  $P$ randomly generated from the Haar measure and resampled for each different setting. Note that, for both $\Sigma_s$ and $\Sigma_d$, the eigenvalues decay slowly to 0, so that no dominating
leading eigenvalue exists.

Under the alternative, for each $p$, $\Sigma$ and each replicate, the mean difference vector $\mu=\mu_1-\mu_2$ is randomly generated from one of the four models:
(1) $\mu\sim N(0,cI_p)$; 
(2) $\mu\sim N(0,c\Sigma)$; 
(3) $\mu\sim N(0,c\Sigma^2)$; and 
(4) $\mu$ is sparse with 5\% randomly selected nonzero entries being either $-c$ or $c$ with probability 1/2 each.
The parameter $c$ is used to control the signal size. The choices in (1)--(4), respectively, represent the cases where $\mu$ is uniform; is slightly tilted towards the
eigenvectors corresponding to large eigenvalues of $\Sigma$; is heavily tilted towards the eigenvectors corresponding to large eigenvalues of $\Sigma$; and is sparse, respectively. 

\begin{figure}
\captionsetup[subfigure]{justification=centering}
\begin{subfigure}[b]{1.2in}
\centering
\includegraphics[width=1.2in,height=1.2in]{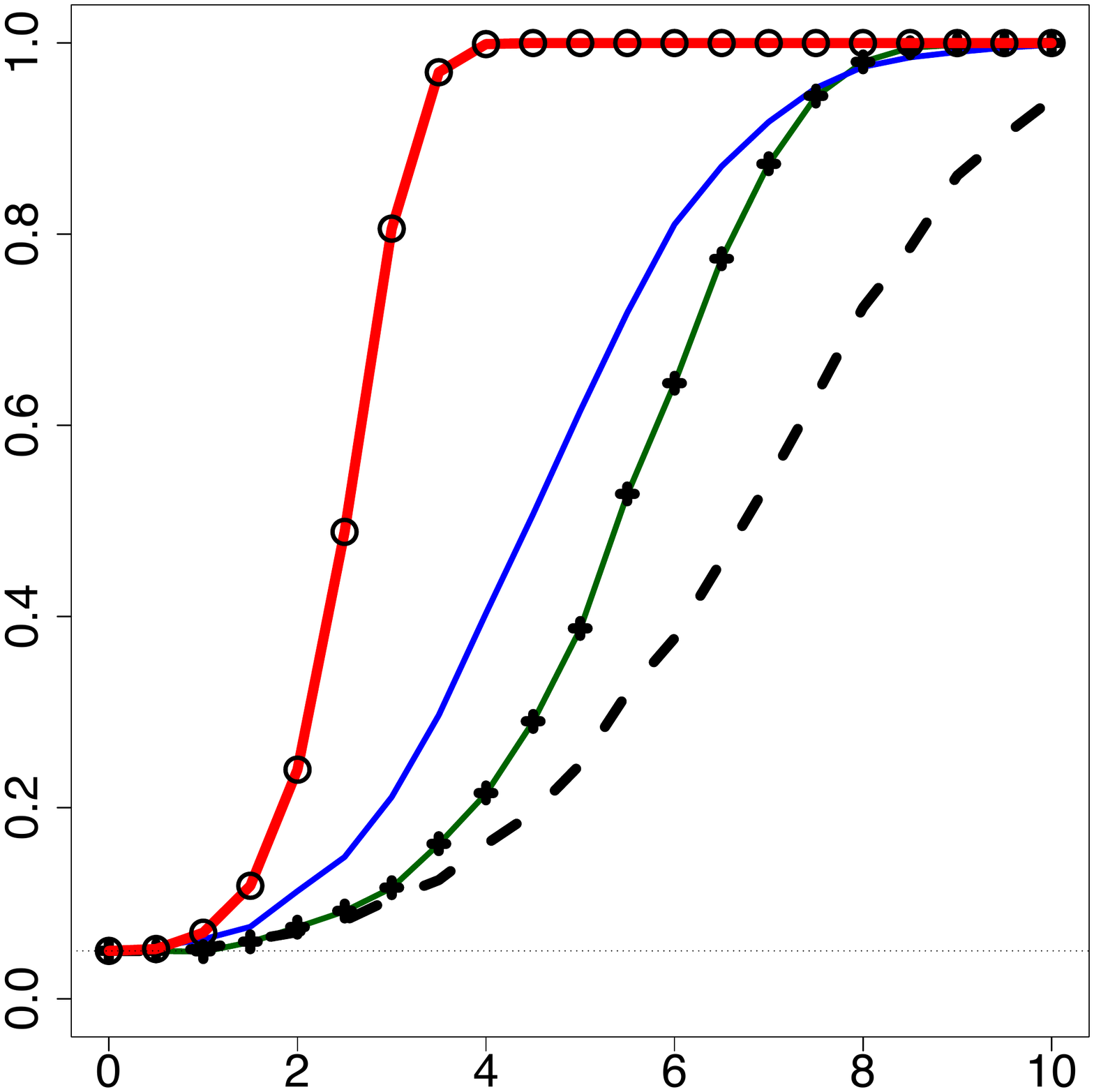}
\subcaption{ $\mu\sim N(0,cI)$ }
\end{subfigure}
\hfill
\begin{subfigure}[b]{1.2in}
\centering
\includegraphics[width=1.2in,height=1.2in]{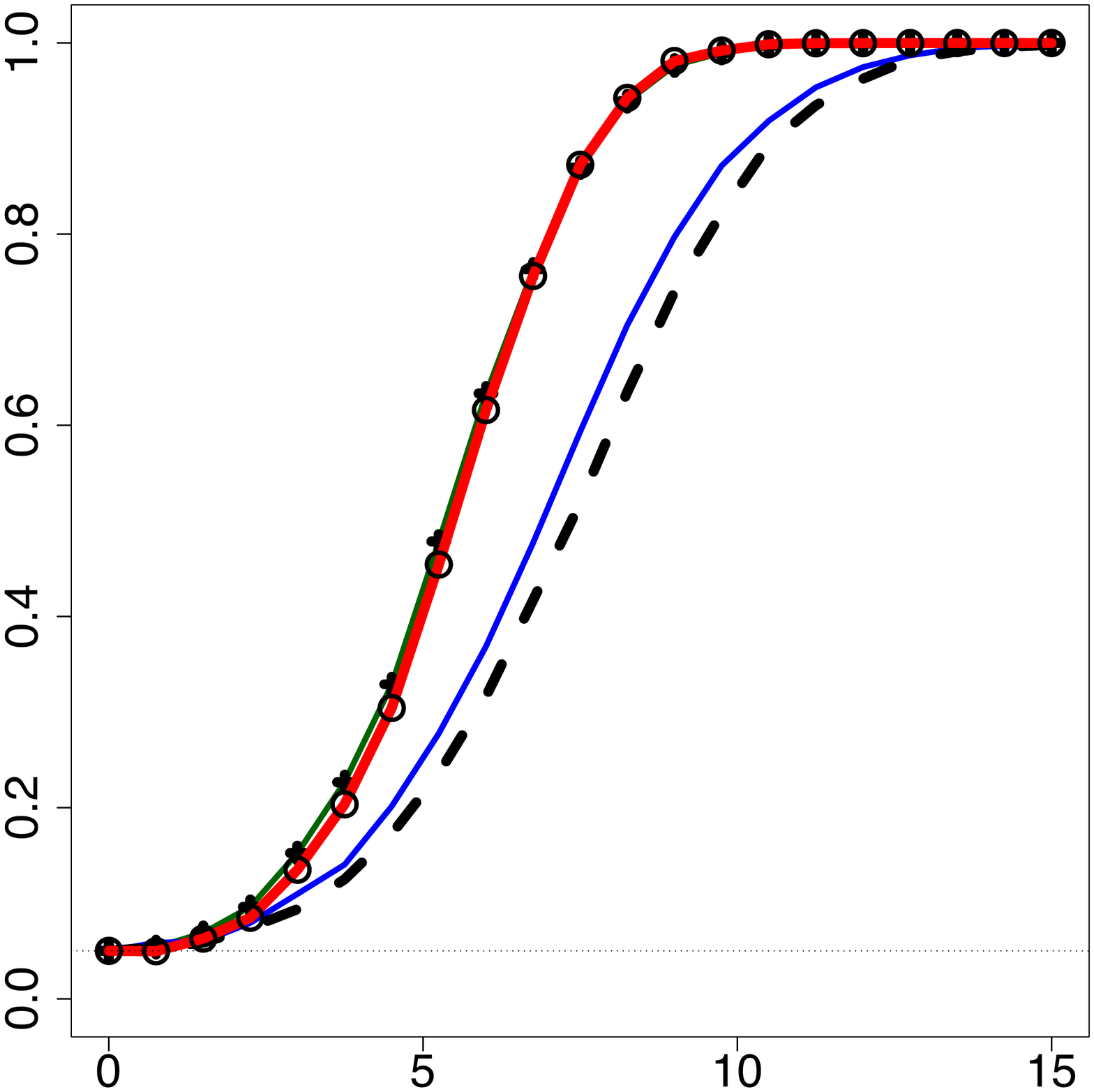}
\subcaption{ $\mu\sim N(0,c\Sigma)$ }
\end{subfigure}
\begin{subfigure}[b]{1.2in}
\centering
\includegraphics[width=1.2in,height=1.2in]{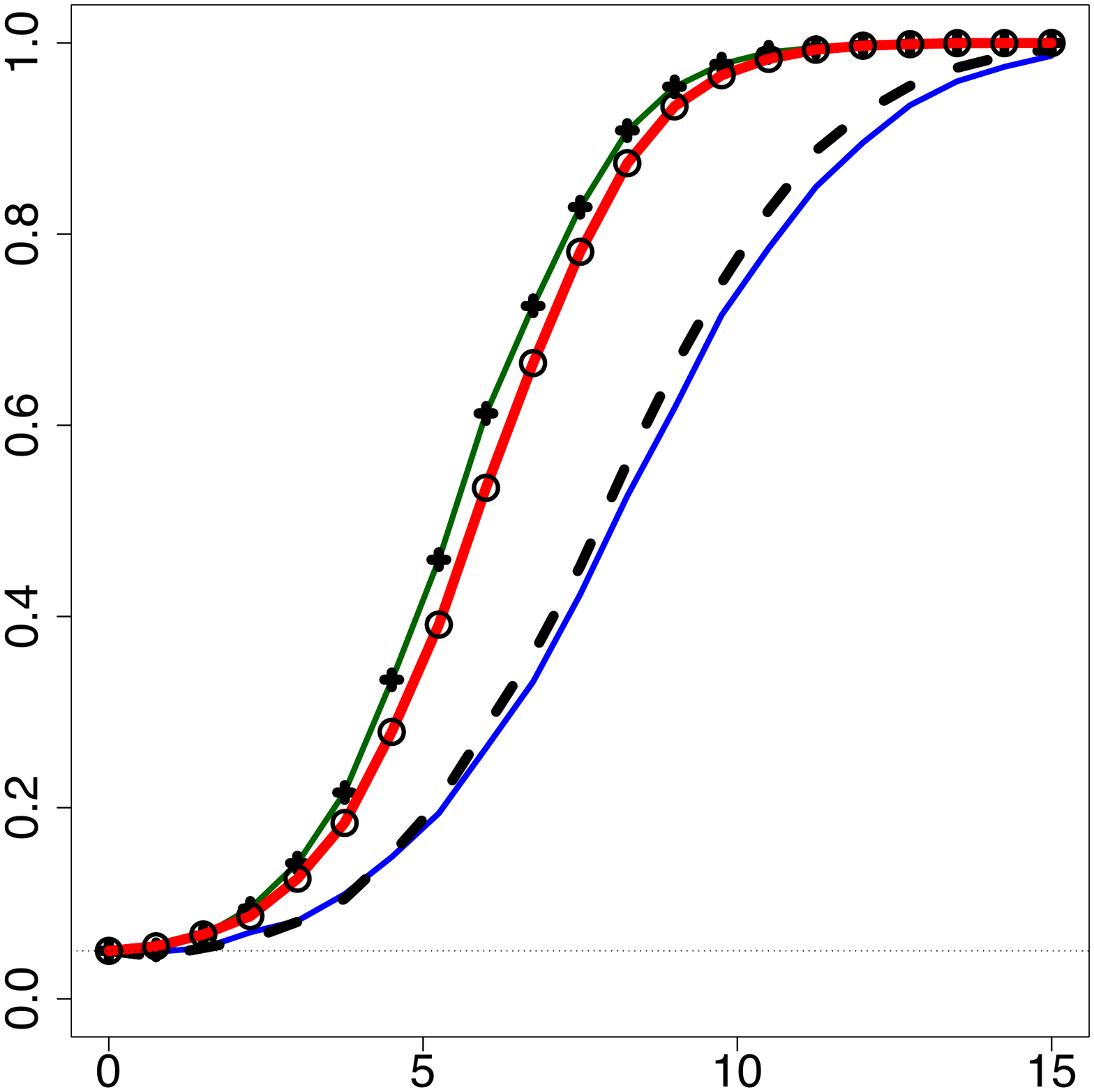}
\subcaption{ $\mu\sim N(0,c\Sigma^2)$ }
\end{subfigure}
\hfill
\begin{subfigure}[b]{1.2in}
\centering
\includegraphics[width=1.2in,height=1.2in]{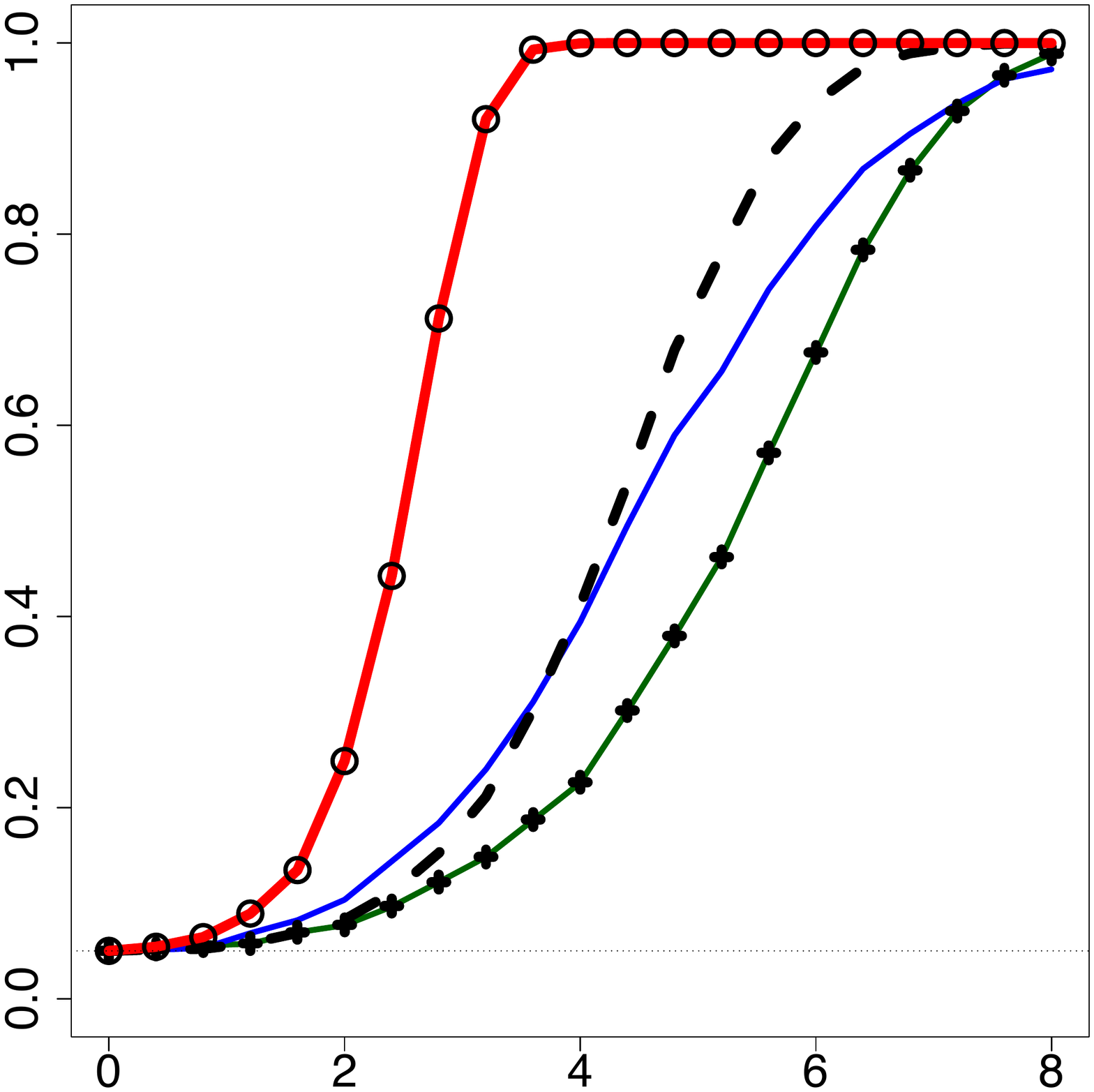}
\subcaption{ sparse $\mu$}
\end{subfigure}
\caption{Size-adjusted empirical power with $X_{ij}\sim N(\cdot,\Sigma)$, $\Sigma=\Sigma_d$ and $p=200$.  $\ARHT_{1/3}$ (solid, red), $\chi^2$ approximation (circle), \citet{BaiS1996} (solid, green) \citet{ChenQ2010} (+), \citet{LopesJW2011} (solid, purple) and \citet{CaiLX2014} with $\Omega$ transform (dashed).}
\label{fig:SizeAdjusted_normal_p_200_Sigma_Dense}
\end{figure}

\begin{figure}
\captionsetup[subfigure]{justification=centering}
\begin{subfigure}[b]{1.2in}
\centering
\includegraphics[width=1.2in,height=1.2in]{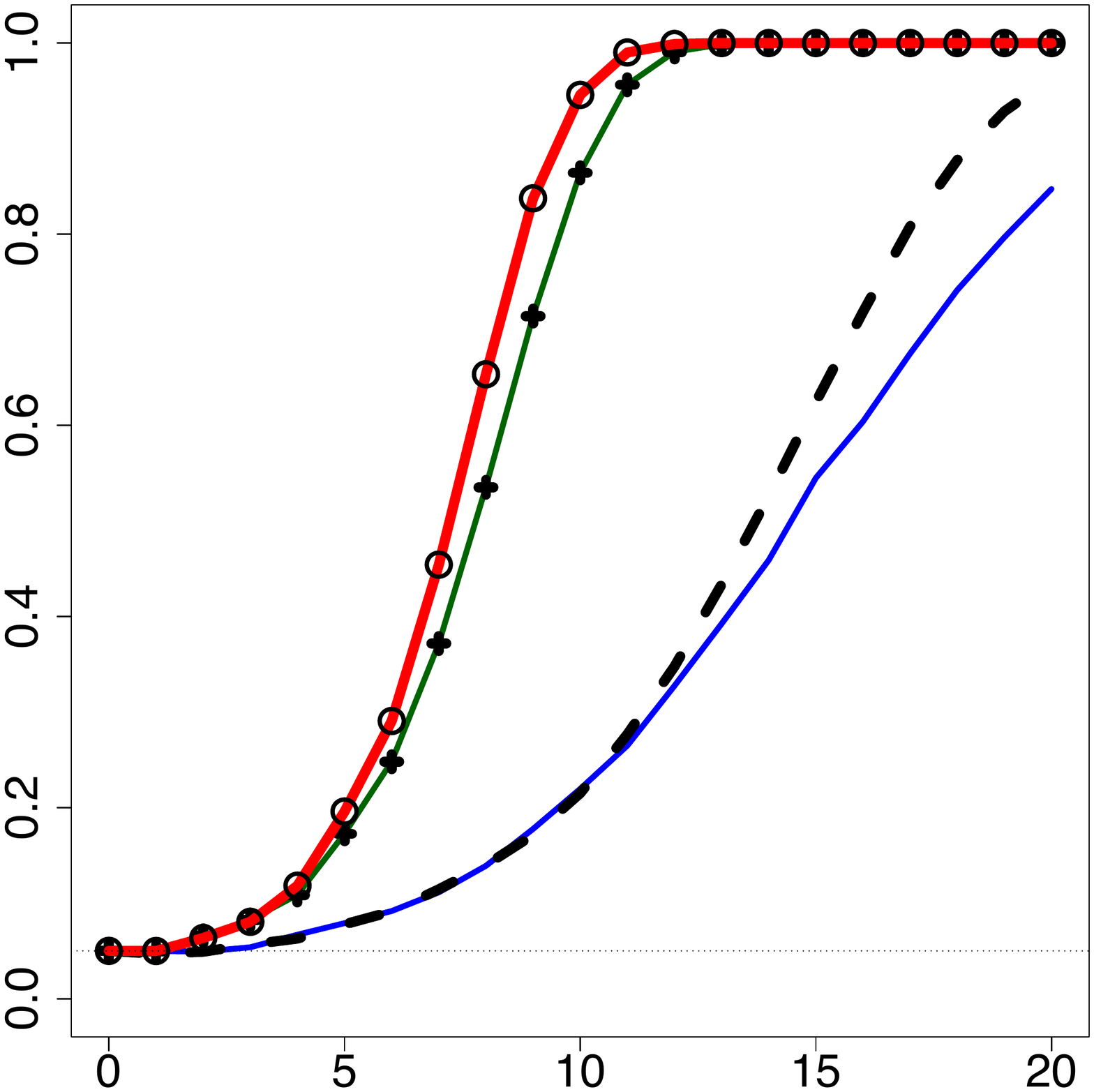}
\subcaption{ $\mu\sim N(0,cI)$ }
\end{subfigure}
\hfill
\begin{subfigure}[b]{1.2in}
\centering
\includegraphics[width=1.2in,height=1.2in]{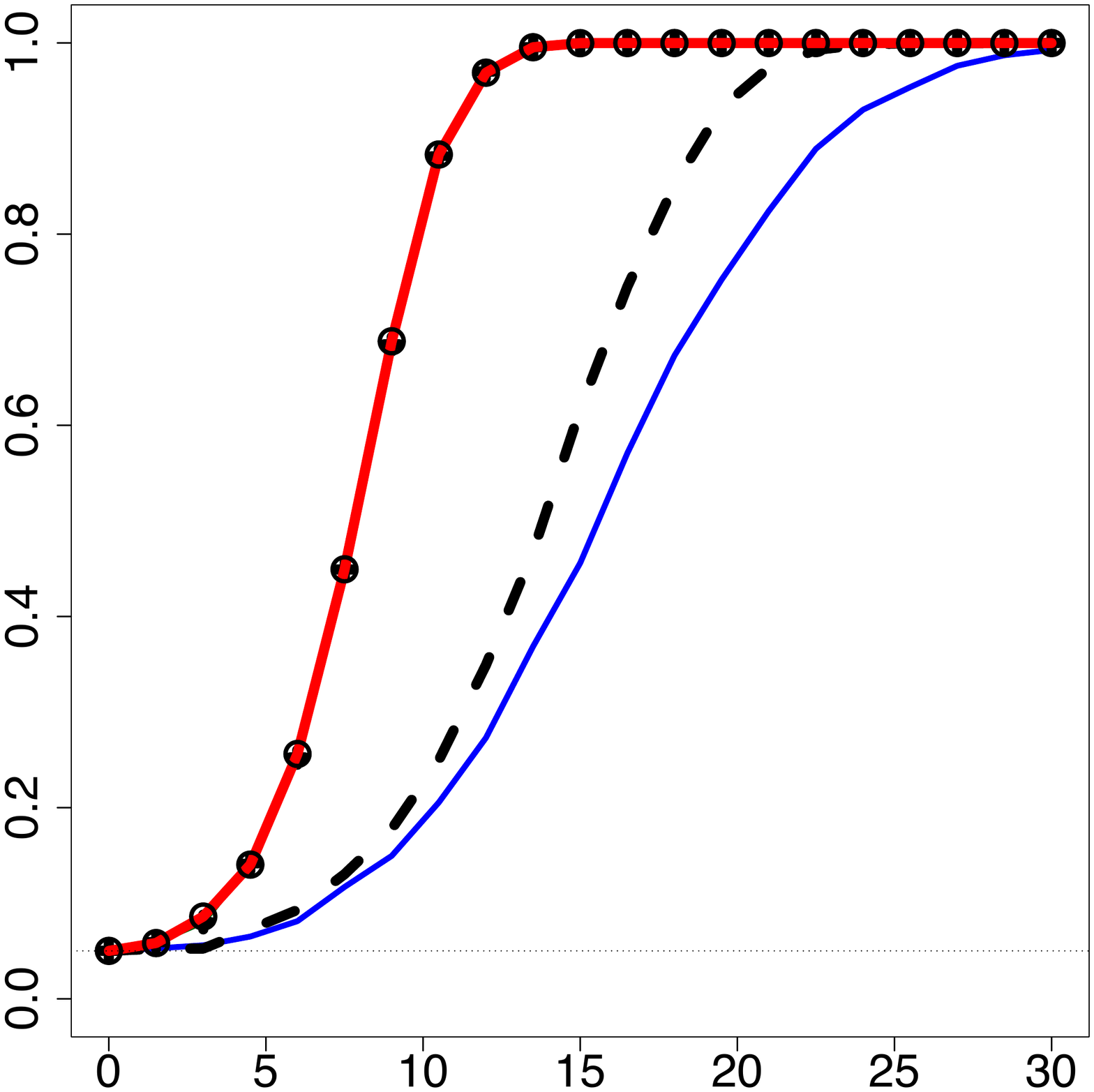}
\subcaption{ $\mu\sim N(0,c\Sigma)$ }
\end{subfigure}
\begin{subfigure}[b]{1.2in}
\centering
\includegraphics[width=1.2in,height=1.2in]{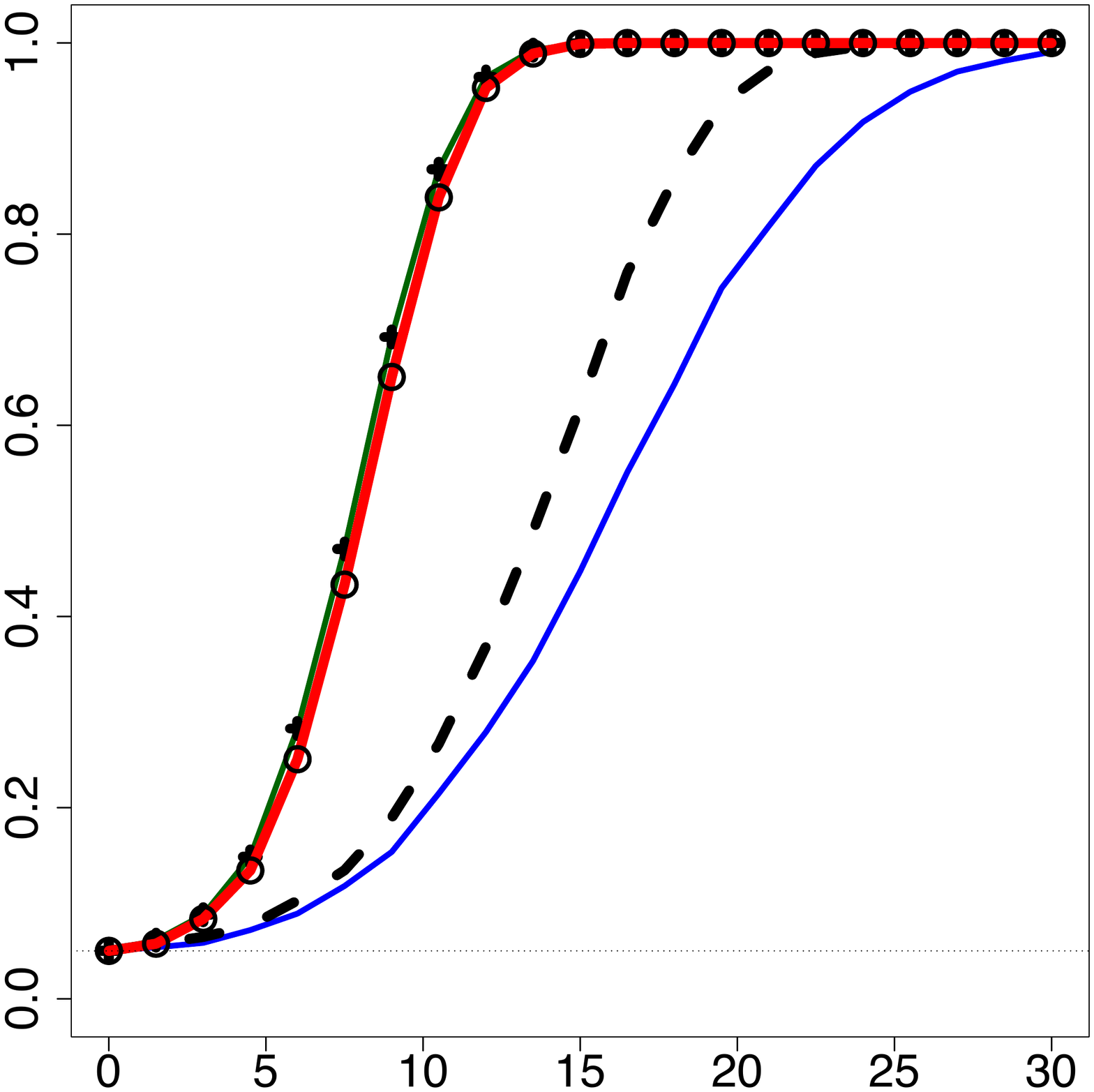}
\subcaption{ $\mu\sim N(0,c\Sigma^2)$ }
\end{subfigure}
\hfill
\begin{subfigure}[b]{1.2in}
\centering
\includegraphics[width=1.2in,height=1.2in]{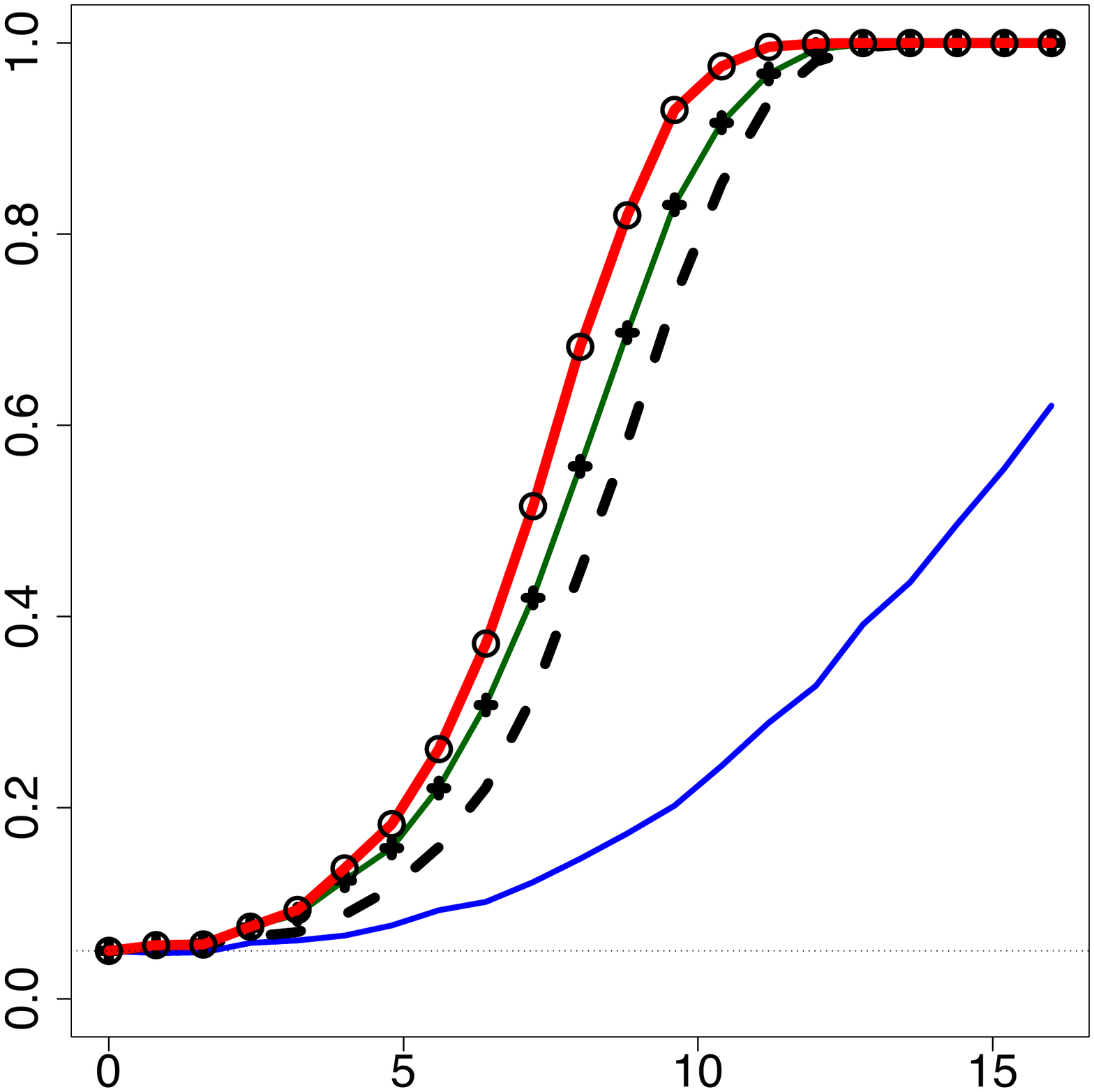}
\subcaption{ sparse $\mu$}
\end{subfigure}
\caption{Same as in Fig. \ref{fig:SizeAdjusted_normal_p_200_Sigma_Dense} but with $p=1000$. }
\label{fig:SizeAdjusted_normal_p_1000_Sigma_Dense}
\end{figure}

\begin{figure}
\captionsetup[subfigure]{justification=centering}
\begin{subfigure}[b]{1.2in}
\centering
\includegraphics[width=1.2in,height=1.2in]{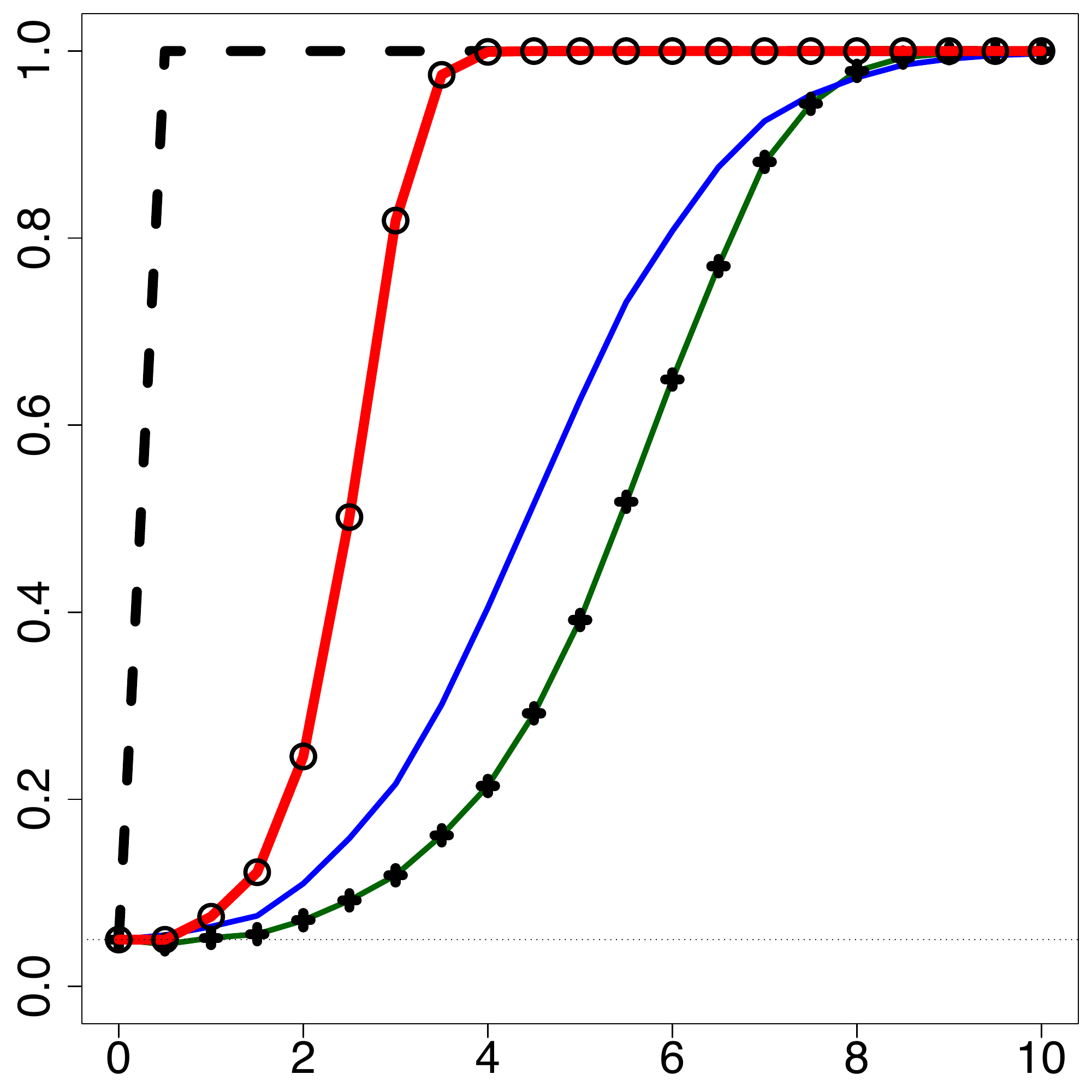}
\subcaption{ $\mu\sim N(0,cI)$ }
\end{subfigure}
\hfill
\begin{subfigure}[b]{1.2in}
\centering
\includegraphics[width=1.2in,height=1.2in]{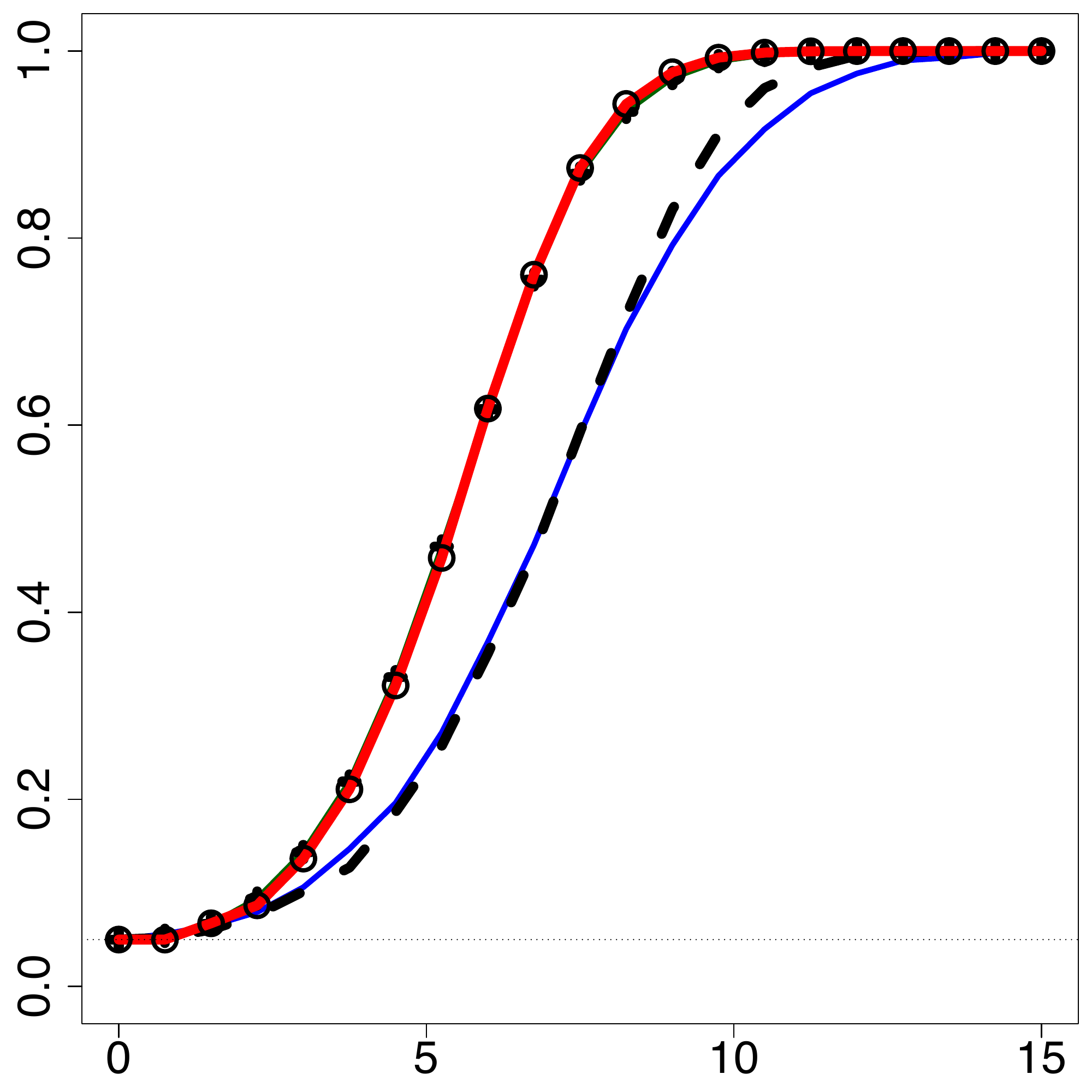}
\subcaption{ $\mu\sim N(0,c\Sigma)$ }
\end{subfigure}
\begin{subfigure}[b]{1.2in}
\centering
\includegraphics[width=1.2in,height=1.2in]{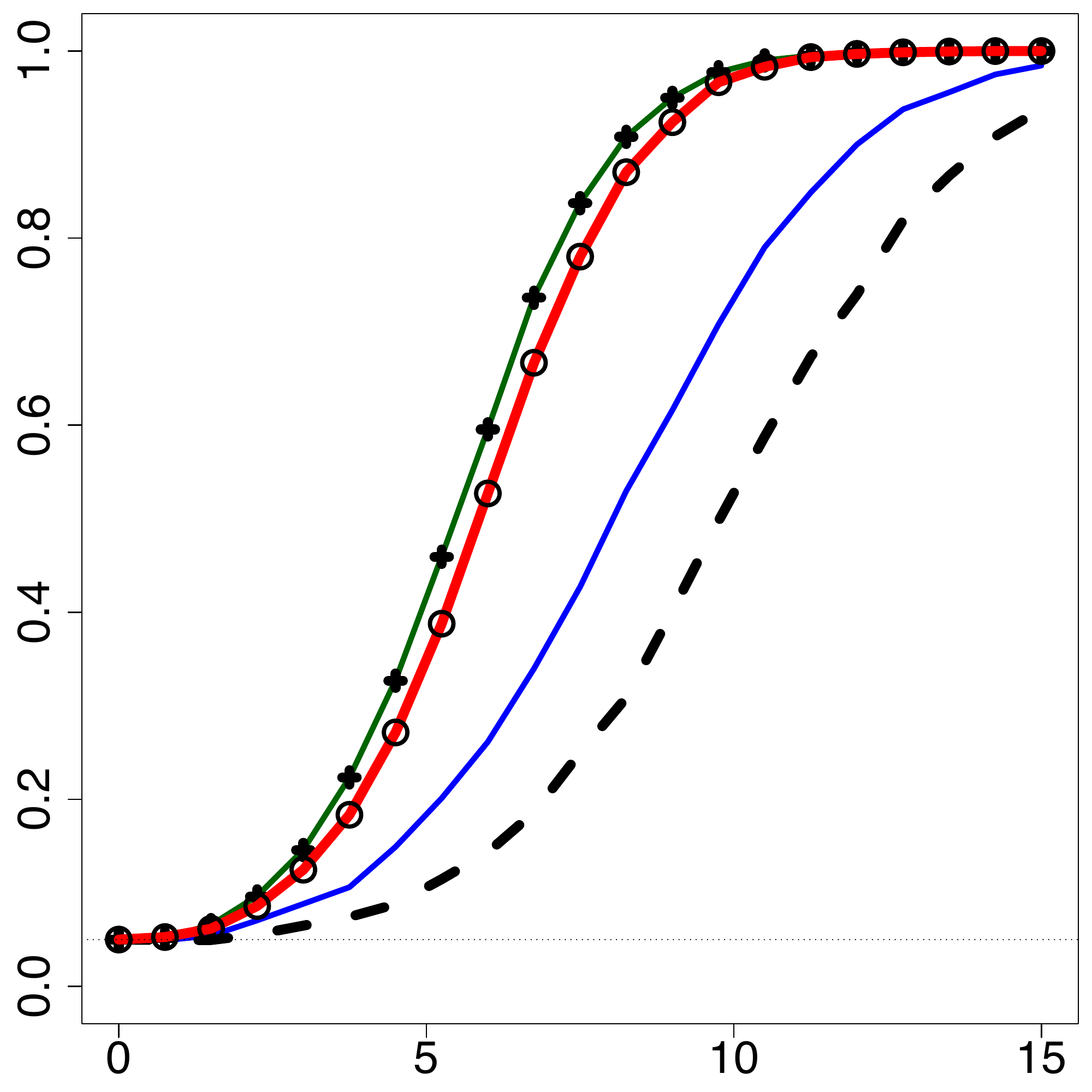}
\subcaption{ $\mu\sim N(0,c\Sigma^2)$ }
\end{subfigure}
\hfill
\begin{subfigure}[b]{1.2in}
\centering
\includegraphics[width=1.2in,height=1.2in]{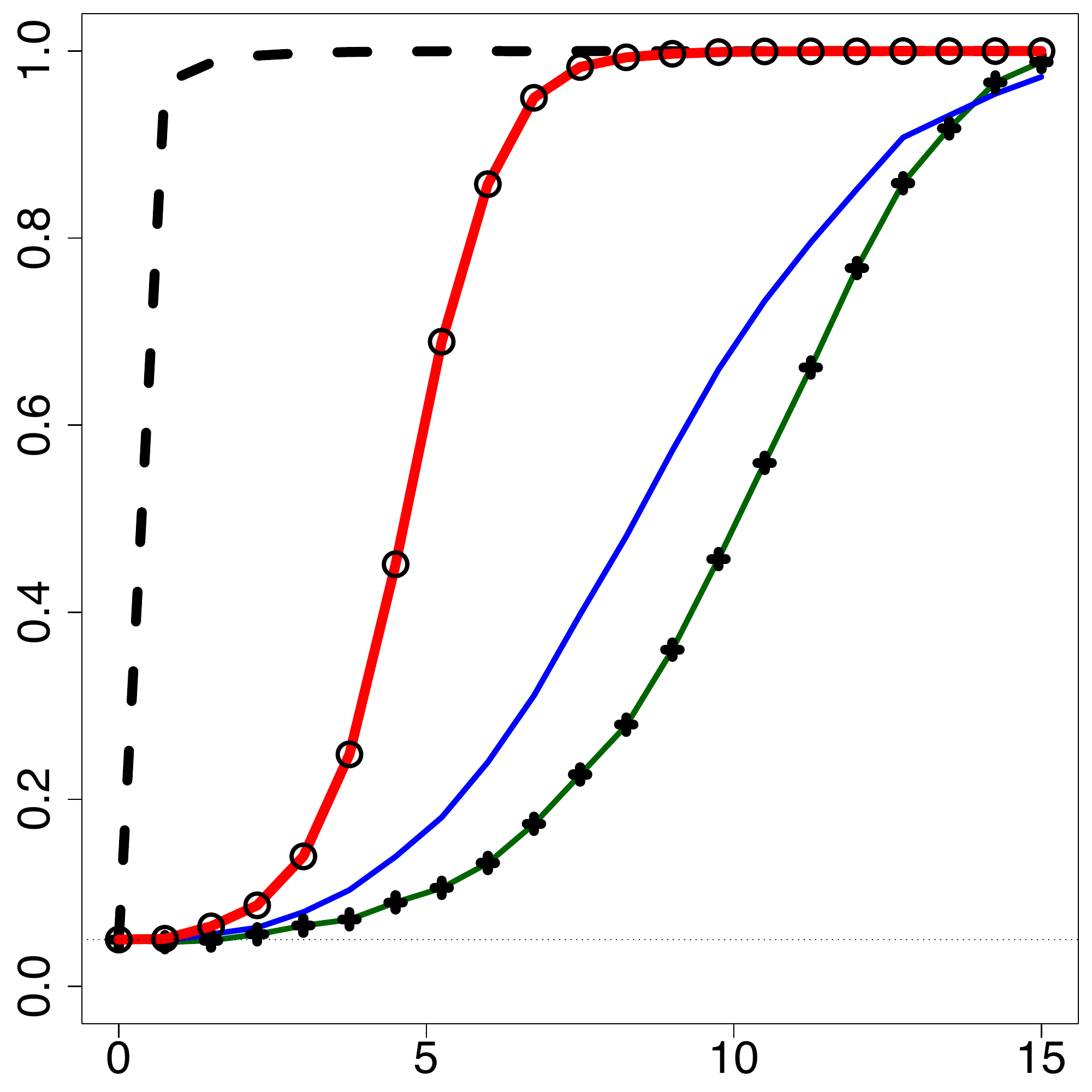}
\subcaption{ sparse $\mu$}
\end{subfigure}
\caption{Size-adjusted empirical power with $X_{ij}\sim N(\cdot,\Sigma)$, $\Sigma=\Sigma_s$ and $p=200$. $\ARHT_{1/3}$ (solid, red), $\chi^2$ approximation (circle), \citet{BaiS1996} (solid, green) \citet{ChenQ2010} (+), \citet{LopesJW2011} (solid, purple) and \citet{CaiLX2014} with  $\Omega$ transform (dashed).}
\label{fig:SizeAdjusted_normal_p_200_Sigma_Sparse}
\end{figure}

All tests are conducted at significance level $\alpha=0.05$. There are two versions for each test: (a) utilizing (approximate) asymptotic cut-off values; and (b) utilizing the size-adjusted cut-off values based on the 
actual null distribution computed by simulations. Only results for the latter case are reported here; the former is in the Supplementary Material. Also, power graphs are given for the Gaussian case only, since power curves for the $
 t_{(4)}$ case are similar (see Supplementary Material). All empirical cut-off values, powers and sizes are calculated based on 10,000 replications. Empirical sizes for the various tests are shown in Table~\ref{tab:true_size}. Empirical power curves versus expected signal strength $(\sqrt{n}\mE[\|\mu\|_2^2])^{1/2}$ are shown in Figures \ref{fig:SizeAdjusted_normal_Sigma_ID}--\ref{fig:SizeAdjusted_normal_p_200_Sigma_Sparse}. Note that, in some of the settings, several of the power curves nearly overlap, creating an occlusion effect. For example, $\mathrm{CLX.}\Omega^{1/2}$ is very similar to $\mathrm{CLX.}\Omega$, therefore only the latter is displayed. For the ease of illustration, power curves corresponding to the recommended $\ARHT_{1/3}$ are plotted as the top layer.

\vspace{.3cm}
\begin{table}[htbp]
\footnotesize
\centering
\caption{Empirical sizes of the various tests at the $\alpha=0.05$ level.}{
\begin{tabular}{l@{\quad}c@{\quad}r@{\qquad}C{26pt}C{26pt}C{26pt}C{26pt}C{26pt}C{26pt}C{26pt}C{26pt}}
\hline
& $\Sigma$ & $p$\phantom{0}  &{\scriptsize $\ARHT$}  & {\scriptsize $\ARHT_{\tiny 1/3}$} & {\scriptsize $\ARHT_{\tiny \chi^2}$} & {\scriptsize BS}      & {\scriptsize CQ} & {\scriptsize RP}   &{\scriptsize CLX.$\Omega^{1/2}$}  &{\scriptsize CLX.$\Omega$}\\
\hline
N(0,1) & ID         & 50  & .0612    & .0447   & .0472   & .0609   & .0481   & .0520   & .0633   & .0637 \\
N(0,1) & ID         & 200 & .0568    & .0473   & .0493   & .0561   & .0508   & .0490   & .0754   & .0757 \\
N(0,1) & ID         & 1000& .0539    & .0491   & .0510   & .0527   & .0517   & .0498   & .1004   & .1004\\[.08cm] 
N(0,1) & $\Sigma_d$ & 50  & .0854    & .0489   & .0606   & .0695   & .0470   & .0485   & .0970   & .1101\\
N(0,1) & $\Sigma_d$ & 200 & .0917    & .0601   & .0705   & .0622   & .0486   & .0503   & .0833   & .0971\\
N(0,1) & $\Sigma_d$ & 1000& .0626    & .0520   & .0347   & .0555   & .0484   & .0510   & .0991   & .0996\\[.08cm]
N(0,1) & $\Sigma_s$ & 50 & .0877     &.0492    & .0603   & .0688   & .0468   & .0508   & .0613   & .0615\\
N(0,1) & $\Sigma_s$ & 200 & .0938    &.0596    & .0707   & .0645   & .0487   & .0503   & .0773   & .0773\\
N(0,1) & $\Sigma_s$ & 1000& .0642    &.0539    & .0347   & .0580   & .0510   & .0486   & .0991   & .0992\\[.08cm]
$t_{(4)}$ & ID      & 50 & .0572     &.0395    & .0414   & .0516   & .0450   & .0477   & .0562   & .0563\\
$t_{(4)}$ & ID      & 200 & .0541    &.0447    & .0456  & .0518  & .0505   & .0504     & .0611   & .0611\\
$t_{(4)}$ & ID      &1000 & .0502    &.0460    & .0443  & .0487  & .0527   & .0493     & .0735   & .0735\\[.08cm]
$t_{(4)}$ & $\Sigma_d$ & 50 & .0836  &.0473    & .0582  & .0659  & .0468  & .0485      & .0815   & .0906\\
$t_{(4)}$ & $\Sigma_d$ & 200 & .0912 &.0582    & .0692  & .0590  & .0484   & .0507     & .0759   &.0838\\
$t_{(4)}$ & $\Sigma_d$ & 1000& .0606 &.0503    & .0313  & .0541  & .0500   & .0494     & .0905   & .0906\\[.08cm]
$t_{(4)}$ & $\Sigma_s$ & 50 & .0812  &.0451    & .0559  & .0634  & .0449   & .0481     & .0512   &.0512\\
$t_{(4)}$ & $\Sigma_s$ & 200 & .0872 &.0551    & .0656  & .0565  & .0469   & .0474     & .0638   &.0638\\
$t_{(4)}$ & $\Sigma_s$ &1000 & .0584 &.0481    & .0246  & .0516  & .0502   & .0495     & .0730   & .0730\\
\hline
\end{tabular}}
\label{tab:true_size}
\end{table}

\begin{figure}
\captionsetup[subfigure]{justification=centering}
\begin{subfigure}[b]{1.2in}
\centering
\includegraphics[width=1.2in,height=1.2in]{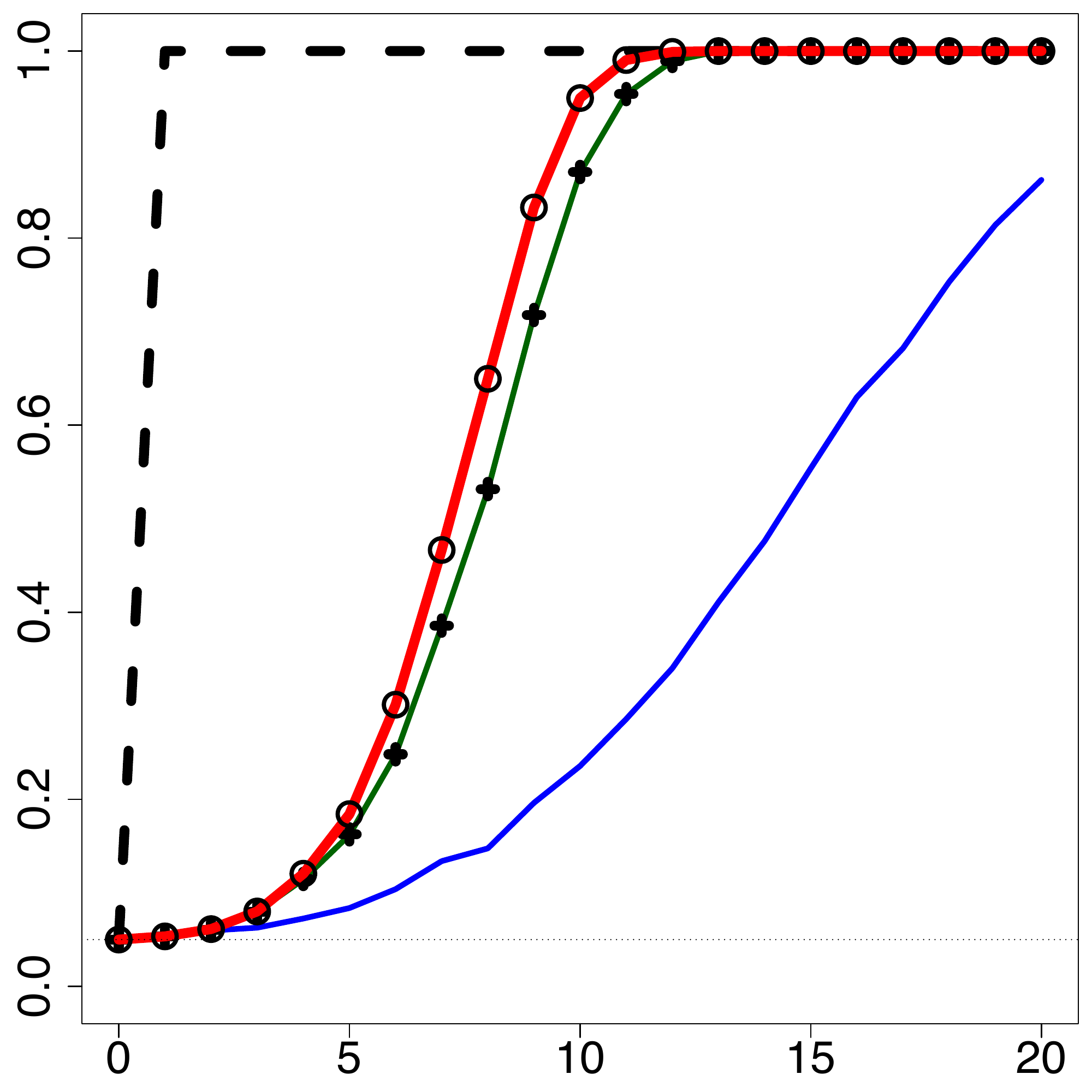}
\subcaption{ $\mu\sim N(0,cI)$ }
\end{subfigure}
\hfill
\begin{subfigure}[b]{1.2in}
\centering
\includegraphics[width=1.2in,height=1.2in]{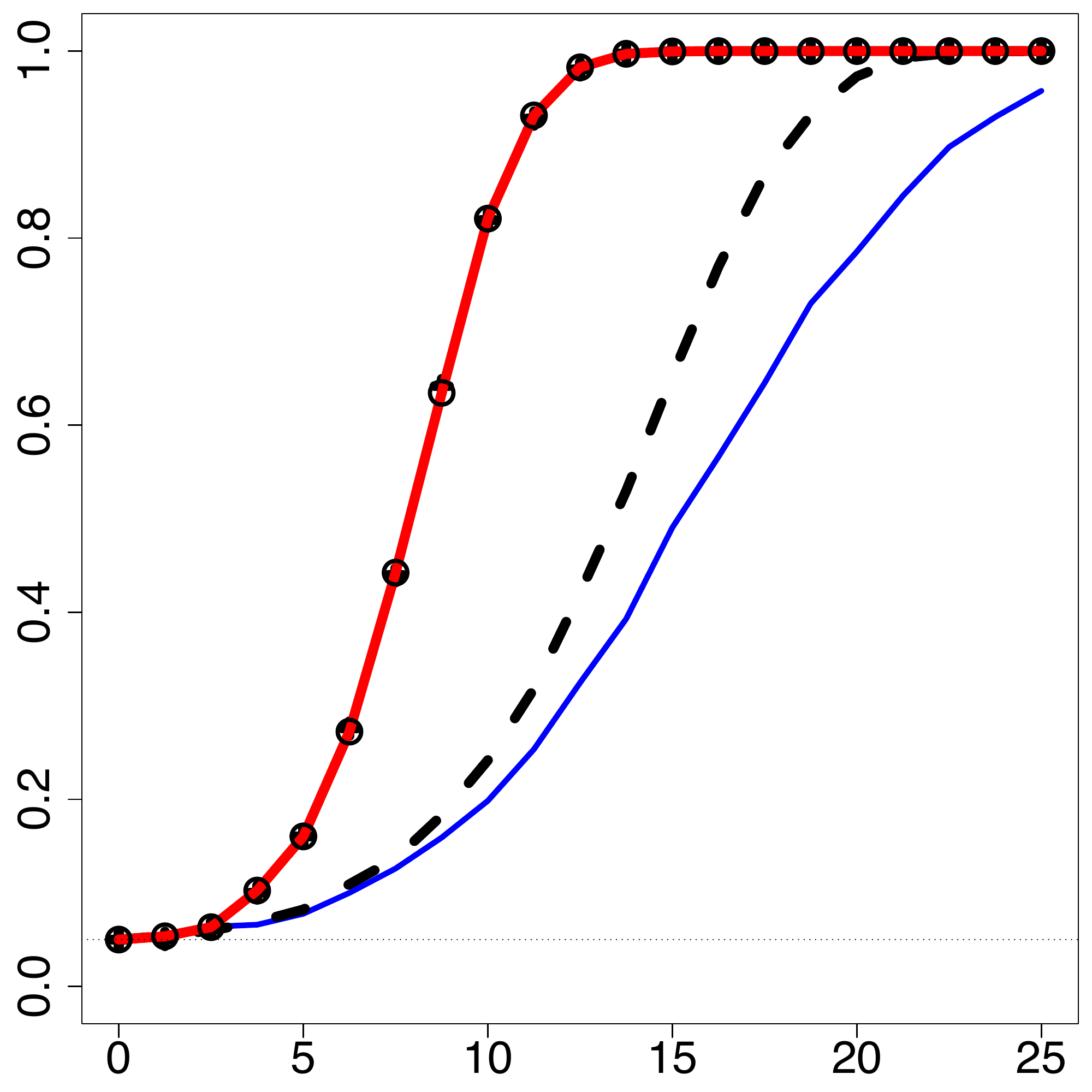}
\subcaption{ $\mu\sim N(0,c\Sigma)$ }
\end{subfigure}
\begin{subfigure}[b]{1.2in}
\centering
\includegraphics[width=1.2in,height=1.2in]{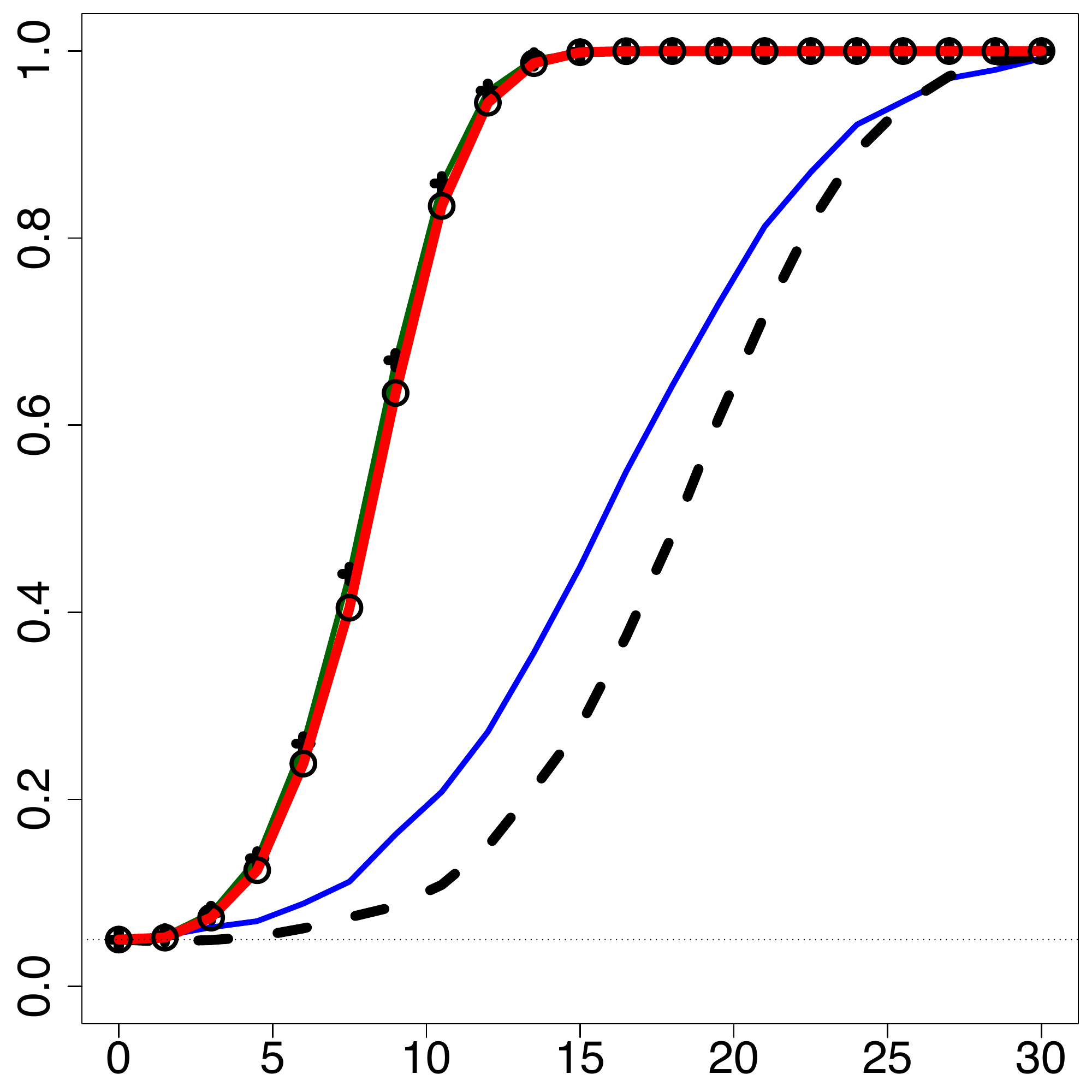}
\subcaption{ $\mu\sim N(0,c\Sigma^2)$ }
\end{subfigure}
\hfill
\begin{subfigure}[b]{1.2in}
\centering
\includegraphics[width=1.2in,height=1.2in]{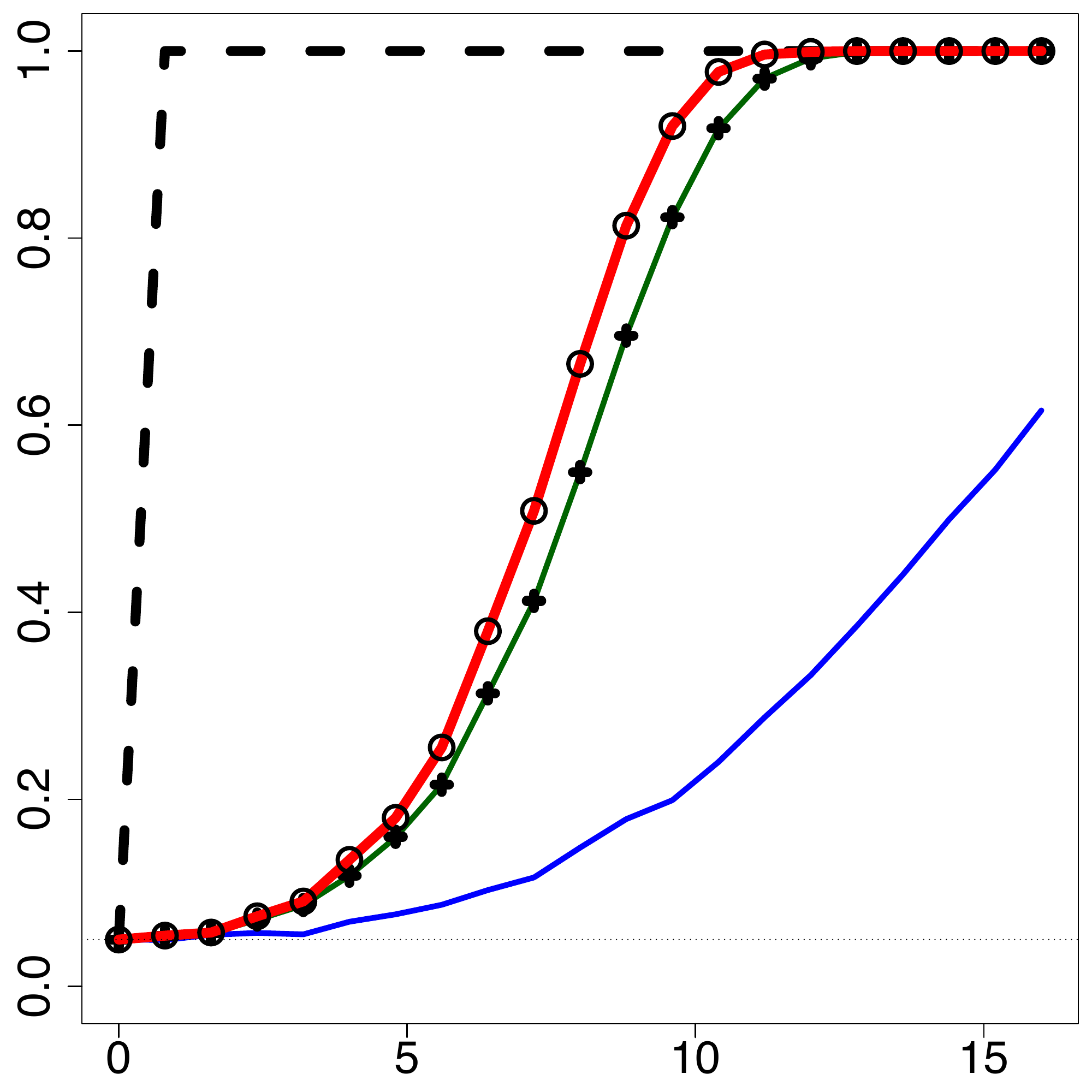}
\subcaption{ sparse $\mu$}
\end{subfigure}
\caption{Same as in Fig. \ref{fig:SizeAdjusted_normal_p_200_Sigma_Sparse} but with $p=1000$.}
\label{fig:SizeAdjusted_normal_p_1000_Sigma_Sparse}
\end{figure}

\subsection{Summary of simulation results}

For each simulation configuration considered in this study, $\ARHT$ or its calibrated versions are as powerful as the procedure(s) with the best performance, except for the cases of sparse or uniform $\mu$ with sparse $\Sigma$ and relatively large $p$ (panels (a) and (d) of Figures \ref{fig:SizeAdjusted_normal_p_200_Sigma_Sparse} and \ref{fig:SizeAdjusted_normal_p_1000_Sigma_Sparse}). This serves as evidence for the robustness of $\ARHT$ procedures with respect to the structures of means under alternatives. The adaptable behavior also sets the proposed methodology apart from its competitors. The following observations are made based on the simulation outcomes.

(1) When the dimension is high and there is no specific structure of $\mu$ and $\Sigma$ that could be exploited, $\ARHT$ tends to outperform the other tests. Tilted alternatives are expected to be detrimental to the performance of both $\ARHT$ and RP. However, $\ARHT$ can be seen as only slightly less powerful than BS and CQ, which yield the best results for this case.

(2) In the case that $\Sigma$ is equal to the identity matrix, the BS procedure is expected to give the best performance, since the test statistic is based on the true covariance matrix. Recalling that BS can be treated as $\RHT(\infty)$, $\ARHT$ is shown to perform as well as BS in corresponding simulations (see Figure \ref{fig:SizeAdjusted_normal_Sigma_ID}). This may be viewed as evidence of the effectiveness of the data-driven tuning parameter selection strategy detailed in Section~\ref{sec:lambda_selection}.

(3) If both mean difference vector $\mu$ and covariance matrix $\Sigma$ are sparse, the three CLX procedures are expected to perform the best. Specifically, the simulations reveal that the sparsity of $\mu$ alone does not guarantee superiority of CLX. This can be seen in the panel (d) of Figures \ref{fig:SizeAdjusted_normal_Sigma_ID}--\ref{fig:SizeAdjusted_normal_p_200_Sigma_Dense}. However, as evidenced in Figures \ref{fig:SizeAdjusted_normal_p_200_Sigma_Sparse} and \ref{fig:SizeAdjusted_normal_p_1000_Sigma_Sparse}, if $\Sigma$ is sparse, then the performance of the CLX procedures is the best when $\mu$ is either uniform or sparse. The $\ARHT$ procedures are less sensitive to the structure imposed on the covariance matrix $\Sigma$ than the CLX procedures, although they are less powerful in sparse settings.

The reason for the excellent performance of CLX for uniform $\mu$ (which is even better than for sparse $\mu$) is that significant signals occur, with high probability due to uniform distribution of signal, at coordinates with very small variance due to their high signal-to-noise ratios. Consequently, 
$l_\infty$-norm based methods, such as the CLX tests, are able to efficiently detect such signals. In contrast, all $l_2$-norm based methods, including $\ARHT$, combine the signals over all coordinates and thus tend to miss such signals since the $l_2$ norm of $\mu$ is relatively small. When $\mu$ is sparse, such a phenomenon also happens but with smaller probability. When $\mu$ is tilted, on the other hand, this phenomenon is unlikely to occur.
Therefore, what is at play is not only sparsity of $\mu$, but also the matching of significant signals with small variances.

The results of this simulation study highlight the robustness or adaptivity of the proposed $\ARHT$ test to various different alternative scenarios and therefore demonstrate its potential usefulness for real world applications. 

\section{Application}\label{sec:application}

Breast cancer is one of the most common cancers with more than 1,300,000 cases and 450,000 deaths worldwide each year. Breast cancer is also a heterogeneous disease, consisting of several subtypes with distinct pathological and clinical characteristics. To better understand the disease mechanisms underlying different breast cancer subtypes, it is of great interest to characterize subtype-specific somatic \textit{copy number alteration (CNA)} patterns, that have been shown to play critical roles in activating oncogenes and in inactivating tumor suppressors during the breast tumor development; see \citep{BreastCGH}. In this section, the proposed $\ARHT$ is applied to a TCGA (The Cancer Genome Atlas) breast cancer data set \citep{TCGA} to detect pathways showing distinct CNA patterns between different breast cancer subtypes.

Level-three segmented DNA copy number (CN) data of breast cancer tumor samples were obtained from the 
TCGA web site. Focus is on a subset of 80 breast tumor samples, which are also subjected to deep protein-profiling 
by CPTAC (Clinical Proteomic Tumor Analysis Consortium) \citep{CPTAC1,CPTAC2,mertins2016proteogenomics}. 
Thus findings from our analysis may lead to further investigations and knowledge generation through the corresponding protein profiles in the future. Specifically, among these 80 samples, 18, 29, and 33 samples belong to the Her2-enriched (Her2), Luminal A (Lum A) and Luminal B (Lum B) subtypes, respectively.
\begin{figure}
\includegraphics[width=4.9in]{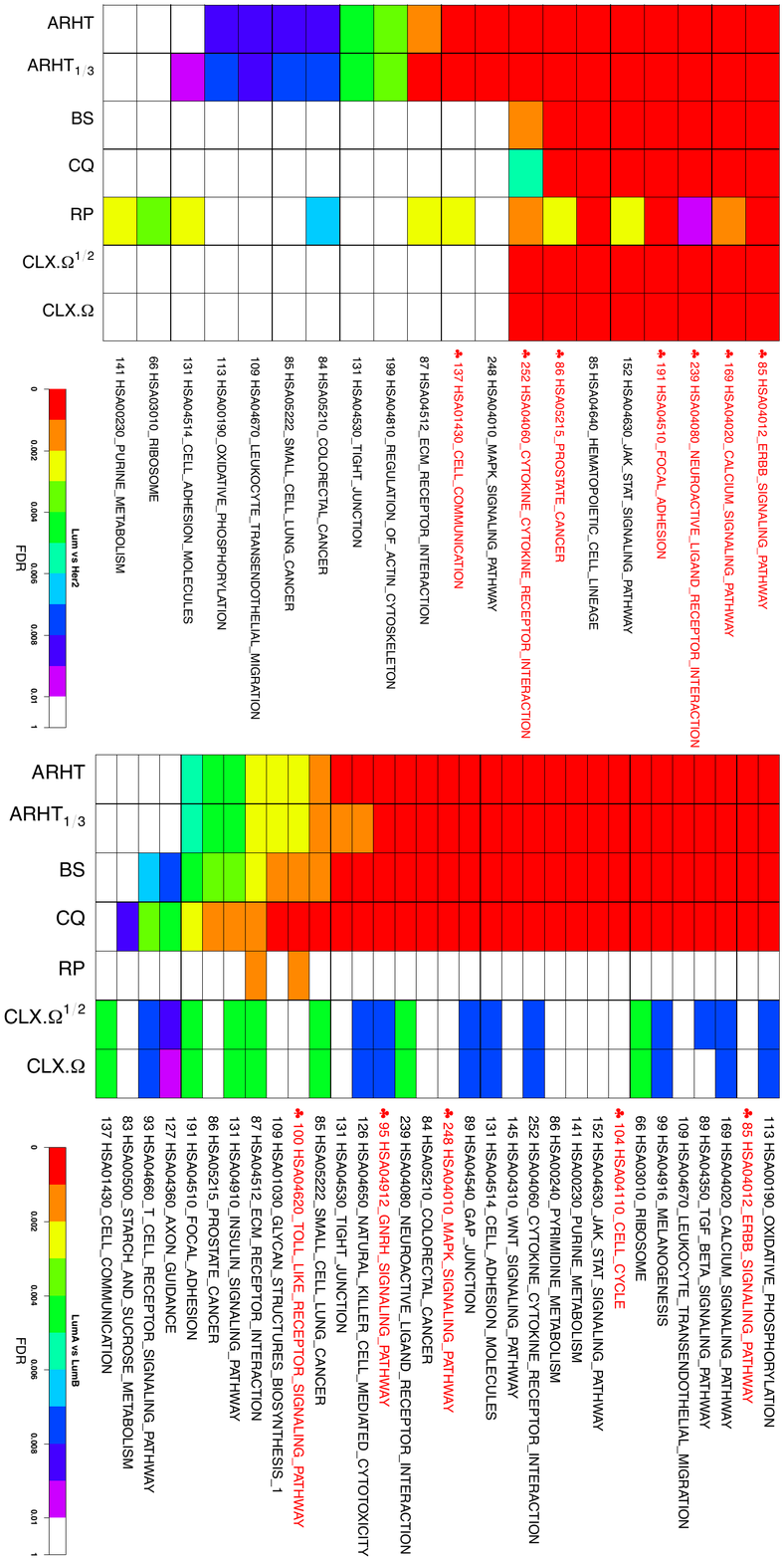}
\caption{Lum vs Her2 (left panel) and Lum A vs Lum B (right panel). Row labels show pathway names and size ($p$), with those known to be significant highlighted by $\clubsuit$ and red color.}
\label{fig:Result_Panel}
\end{figure}

For the selected samples, first gene-level copy number estimates are derived based on the segmented CN profiles. Q-Q plots, provided in the Supplementary Material, suggest that the observations have heavier tails than normal distributions. To better illustrate the comparative performance of the proposed methods under high dimensions, 
consider the 36 largest KEGG pathways. The
number of genes in these pathways ranging from 66 to 252, so that $p/n$ varies between 0.75 and 3.5. For each pathway, interest is in testing whether genes in the pathway showed different copy number alterations between Lum (Lum A plus Lum B) vs.\ Her2, or Lum A vs.\ Lum B. These led to a total of 72 two-sample tests.

All testing methods discussed in the simulation studies were applied to this data set, except for $\ARHT_{\chi^2}$. 
The null distribution and the $p$-value for each method, were generated based on 100,000 permutations, instead of applying the asymptotic theory, though the asymptotic and permutation-based cut-offs are similar for $\ARHT_{1/3}$. Also, to control the family-wise error rate, the $p$-values are further adjusted by FDR \citep{FDR}, and FDR-adjusted $p$-values below 0.01 indicate departure from null.
For the Lum vs Her2 comparison, $\ARHT$ yielded the largest number of significant pathways followed by RP, while all other methods have similar behaviors with about half the detection rate of $\ARHT$ and RP. For the Lum A vs Lum B comparison, the $\ARHT$ results are similar to those of BS and CQ, giving the largest number of significant pathways. On the other hand, in this case, RP only detected two while the three CLX methods did not detect any significant pathway. 

One unique characteristic of Her2 subtype tumors is the amplification of gene ERBB2 and its neighboring genes in cytoband 17q12, including MED1, STARD3 and others. There are 7 pathways containing at least one of these genes. These pathways, whose annotations were colored in red in Figure \ref{fig:Result_Panel}, can serve as positive controls in the Her2 vs Lum comparison \citep{LumHer2}. Moreover, it has been shown that gene MAP3K1 and MAP2K4 have different CN loss activities in Lum A and Lum B tumors \citep{MAP2K1}. In addition, proliferation genes such as CCNB1, MKI67 and MYBL2 are more highly expressed in Lum B compared to Lum A, as shown in \cite{CCNB1}. Thus, the pathways containing these genes can be viewed as positive controls in the Lum A vs Lum B comparison analysis. As an illustrative reference, in Table \ref{table:Significant_performance}, the performance of different procedures is summarized in terms of detecting the pathways known to have different CN alterations between subtypes, when FDR is controlled at 0.01. Interestingly, only the three $\ARHT$ procedures successfully detected all these pathways of positive
controls, suggesting a superior power of $\ARHT$ procedures over the competitors. BS and CQ appeared to be
the second best methods. 

\begin{table}
\centering
\def~{\hphantom{0}}
\caption{Comparative performance on known significant pathways
  (at FDR level $0.01$).}{
\begin{tabular}{l|c|c}
\hline\hline
           & ~Lum~ vs ~Her2~ & ~Lum A~ vs ~Lum B~\\
           \hline
$\ARHT$~        &  7/7        &    5/5        \\
 \hline
$\ARHT_{1/3}$~  &  7/7        &    5/5        \\
 \hline
BS~         &  6/7        &    5/5        \\
 \hline
CQ~         &  6/7        &    5/5        \\
 \hline
RP~         &  7/7        &    1/5        \\
 \hline
CLX.$\Omega^{1/2}$~ & 6/7 &    1/5        \\
 \hline
CLX.$\Omega$~       & 6/7 &    1/5        \\
 \hline \hline
\end{tabular}}
\label{table:Significant_performance}
\end{table}

In summary, for this data, only $\ARHT$ consistently makes correct decisions on pathways known to be significant,
while the other methods perform adequately for at most one of the comparisons -- either Lum vs.\ Her2 or Lum A vs.\ Lum B. This provides further evidence in support of the power and robustness of $\ARHT$.

\section{Discussion}\label{sec:discussion}

In this paper, a powerful and computationally tractable procedure for testing equality of mean vectors between two populations was presented that is based on a composite ridge-type regularization of Hotelling's $T^2$ statistics. Techniques from random matrix theory were used to derive the asymptotic null distribution under a regime where the dimension is comparable to the sample sizes.  Extensive simulations were conducted to show that the proposed test has excellent power for a wide class of alternatives and is fairly robust to the structure of the covariance matrix as well as the distribution of the observations. Practical advantages of the proposed test were illustrated in the context of a breast cancer data analysis where the goal was to detect pathways with different DNA copy number alteration patterns between cancer subtypes.

There are several future research directions that to pursue. On the technical side, aim could be on relaxing the distributional assumptions on the observations further, only requiring the existence of a certain number of moments. 
On the methodological front, aim could be on the extension of the framework to tests for mean difference 
under possibly unequal variances, and to deal with the MANOVA problem in high-dimensional settings. 
Another potentially interesting direction is to combine the proposed methodology with a variable screening strategy so that the test can be adapted to ultra-high dimensional settings. 

\section{Proofs of the main results}
\label{sec:proofs}


In this section, we provide the necessary technical support for the proposed methodology under the class of sub-Gaussian distributions $\mathcal{L}(c_1,c_2)$ introduced in Section \ref{sec:non_gaussianity}. The technical details consist of the following four parts: 
(i) proof of asymptotic normality;
(ii) proof of Theorem \ref{thm:local_power} and Theorem \ref{thm:local_power_probabilistic}; 
(iii) proof of Theorem \ref{thm:converg_lambda}; and
(iv) proof of Theorem \ref{thm:convergence_process}.

The crucial difference between Gaussianity and non-Gaussianity is that in the Gaussian case, the sample covariance matrix $S_n$ is independent of the sample means and can be written as sum of independent random elements. Indeed, under Gaussianity, $S_n=\sum_{i=1}^{n-2}\Sigma^{1/2}_pY_iY_i^T\Sigma^{1/2}_p$ with $Y_j\sim\mathcal{N}(0,(n-2)^{-1}I_p)$ is independent of the $\bar{X}_i$'s, with the latter normally distributed. However, in non-Gaussian settings, due to lack of independence between $S_n$ and $\bar{X}_i$'s, their mutual correlation has to be disentangled carefully. 

For this analysis, following common practice in random matrix theory, we use an un-centered version of the sample covariance, defined as 
\[
\tS=\frac{1}{n}\sum_{j=1}^{n}X_{ij}X_{ij}^T.
\]
Note that
\begin{equation*}
S_n = \frac{n}{n-2} \tS - \frac{n_1}{n-2} \bar{X}_1\bar{X}_1^T - \frac{n_2}{n-2} \bar{X}_2\bar{X}_2^T.
\end{equation*}
The statistic $(\bar{X}_1-\bar{X}_2)^T(S_n+\lambda I_p)^{-1}(\bar{X}_1-\bar{X}_2)$ changes nontrivially if $S_n$ is replaced with $\tS$. It will be shown in the following proofs how to manipulate their difference. 
Recall the following definitions:
\begin{align*}
R_n(z)
&=(S_n-z I_p)^{-1},\\[0.2cm]
\hat\phi_1 
&=p^{-1}\tr(S_n),\\[0.2cm]
\mF(-\lambda)
&=p^{-1}\tr\{R_n(-\lambda)\},\\[0.2cm]
\hat\Theta_1(\lambda,\gamma_n)
&= \frac{1- \lambda \mF(-\lambda)}{1-\gamma_n\{1 - \lambda \mF(-\lambda)\}}, \\[.2cm]
\hat\Theta_2(\lambda,\gamma_n)
&=\frac{1- \lambda \mF(-\lambda)}{[1-\gamma_n\{1 - \lambda \mF(-\lambda)\}]^3} -
\lambda\frac{\{\mF(-\lambda) - \lambda \mF'(-\lambda)\}}{[1-\gamma_n\{1 - \lambda \mF(-\lambda)\}]^4}.
\end{align*}
For the sake of brevity, $S_n$ is replaced with $\tS$ in all these quantities and proofs are provided, even in the Gaussian case, using the thus modified versions. Because 
\begin{align*}
&\left|p^{-1}\tr(S_n)-p^{-1}\tr(\tS)\right|= O_p(1/p),\\
&\left|p^{-1}\tr\{(S_n+\lambda I_p)^{-k}\}- p^{-1}\tr\{(\tS+\lambda I_p)^{-k}\}\right|\leq \frac{2k}{\lambda^k p},
\end{align*}
all the derivations all results put forward in the rest of this section will also hold for the original quantities.
The argument for the first relation is straightforward and the second argument is deduced from Lemma \ref{lemma:bounded_diff}. 
To lighten notation, $\hat\phi_1$, $R_n(z)$, $m_{F_{n,p}}(-\lambda)$, $\hat\Theta_1(\lambda,\gamma_n)$, $\hat\Theta_2(\lambda,\gamma_n)$, etc., are used to denote their counterparts after the replacement of $S_n$ by $\tS$.

As mentioned above, the proposed statistic and other quadratic terms involving $S_n$ will change significantly after the redefinition of $S_n$. Define
\begin{equation}\label{eq:U_i_iprime_def}
U_{ii'}(\lambda) 
= \bar{X}_i^T(\tS+\lambda I_p)^{-1}\bar{X}_{i'}, 
\qquad i,i'=1,2. 
\end{equation}
The Woodbury matrix identity gives
\begin{align}
\label{eq:S_ts}
&\frac{n}{n-2}\bigg(S_n +\frac{n}{n-2}\lambda I_p\bigg)^{-1} \\ 
&=(\tS+\lambda I_p)^{-1} + 
(\tS+\lambda I_p)^{-1}(\bar{X}_1,\bar{X}_2)\mathbb{H}^{-1} 
 \Big(
\begin{matrix}
\bar{X}_1^T \\[2pt]
\bar{X}_2^T
\end{matrix}
 \Big) (\tS+\lambda I_p)^{-1}, 
\nonumber
\end{align}
where 
\[
\mathbb{H}= \Big(
\begin{matrix}
{n}n^{-1}_1 &0\\[2pt]
0     &{n}n_2^{-1}
\end{matrix}\Big) - 
\Big(\begin{matrix} U_{11}(\lambda)& U_{12}(\lambda) \\ U_{21}(\lambda) & U_{22}(\lambda) \end{matrix} \Big)
.
\]
Therefore, 
\begin{equation}
\label{eq:RHT_U}
\begin{split}
\frac{n}{n-2}\RHT(\frac{n}{n-2}\lambda)=& \frac{n_1n_2}{n_1+n_2}\Big[ (U_{11}(\lambda) + U_{22}(\lambda) -2 U_{12}(\lambda)) \\
& ~~ + ~~\Big(
\begin{matrix} U_{11}(\lambda) -U_{12}(\lambda) \\[2pt]
U_{12}(\lambda) -U_{22}(\lambda) \end{matrix}
\Big)^T  \mathbb{H}^{-1} \Big(
\begin{matrix} U_{11}(\lambda) -U_{12}(\lambda) \\[2pt]
 U_{12}(\lambda) -U_{22}(\lambda) \end{matrix}
\Big)  \Big]. 
\end{split}
\end{equation}

\subsection{Proof of asymptotic normality under sub-Gaussianity}
\label{subsec:asy_normality}

It follows from \eqref{eq:RHT_U} that $\RHT(n(n-2)^{-1}\lambda)$ can be expressed as a differentiable function of $U_{11}(\lambda)$, $U_{12}(\lambda)$ and $U_{22}(\lambda)$. Hence, the joint asymptotic normality of the latter implies the asymptotic normality of the former. Therefore, define an arbitrary linear combination,
$$\bar{R}(\lambda)=n^{1/2}[l_{11}U_{11}(\lambda)+l_{12}U_{12}(\lambda)+l_{22}U_{22}(\lambda)]$$
for any $l_{11}, l_{12}, l_{22}\in\mR$. It suffices to show that $\bar{R}(\lambda)$ is asymptoticly normal. 

To this end, we use Theorem \ref{thm:Chatterjee2009}. A key component of the proof is to establish the asymptotic orders of $\varrho_0(\bar{R})$, $\varrho_1(\bar{R})$ and $\varrho_2(\bar{R})$ and also $\var(\bar{R})$. Since the gradient and Hessian of $\bar{R}(\lambda)$ are linear functions of those of $n^{1/2}U_{11}(\lambda)$, $n^{1/2}U_{12}(\lambda)$ and $n^{1/2}U_{22}(\lambda)$, it suffices
to derive asymptotic orders of the functions $\varrho_0$, $\varrho_1$ and $\varrho_2$ with  $n^{1/2}U_{11}$, $n^{1/2}U_{12}$ and $n^{1/2}U_{22}$ as arguments, then combining them through the Cauchy--Schwarz inequality. In the rest of the proof, only the asymptotic 
order of $\varrho_0({p}^{1/2}U_{11})$, $\varrho_1({p}^{1/2}U_{11})$ and $\varrho_2({p}^{1/2}U_{11})$ is derived as similar arguments also work for $U_{12}$ and $U_{22}$. 

\begin{proposition}
	\label{proposition:k_0}
	Under the assumptions of Theorem \ref{general},
	$\varrho_0(\sqrt{n}U_{11})=o(1)$,
\end{proposition}
\begin{proposition}
	\label{proposition:k_1}
	Under the assumptions of Theorem \ref{general},
	$\varrho_1(\sqrt{n}U_{11})=o({n}^{1/2})$.
\end{proposition}
\begin{proposition}
	\label{proposition:k_2}
	Under the assumptions of Theorem \ref{general},
	$\varrho_2(\sqrt{n}U_{11})=O({n^{-1/2}})$.
\end{proposition}
\begin{proposition}
\label{proposition:asym_mean_variance}
Under the assumptions of Theorem \ref{general},
\begin{align*}
\label{eq:mean_variance}
&\mE \bar{R}(\lambda) = \frac{(l_{11}/k + l_{22}/(1-\kappa))\gamma \Theta_1(\lambda,\gamma)}{1+\gamma\Theta_1(\lambda,\gamma)} +o(1),\\[10pt]
&\mathrm{Var}(\bar{R}(\lambda)) = \frac{[2l_{11}^2/\kappa^2 + l_{12}^2/(\kappa-\kappa^2) + 2l_{22}^2/(1-\kappa)^2]\gamma^2 \Theta_2(\lambda,\gamma)}{(1+\gamma\Theta_1(\lambda,\gamma))^4} + o(1),\\[10pt]
&\mathrm{Cov}(\bar{R}(\lambda),\bar{R}(\lambda')) \\[10pt]
&\qquad= \frac{[2l_{11}^2/\kappa^2 + l_{12}^2/(\kappa-\kappa^2) + 2l_{22}^2/(1-\kappa)^2]\gamma^2 \Theta_3(\lambda,\lambda',\gamma)}{(1+\gamma\Theta_1(\lambda,\gamma))^2(1+\gamma\Theta_1(\lambda',\gamma))^2} + o(1),
\end{align*}
where for $\lambda\neq \lambda'$,
\[\Theta_3(\lambda,\lambda',\gamma) =(1+\gamma\Theta_1(\lambda,\gamma))(1+\gamma\Theta_1(\lambda',\gamma)) \frac{(\lambda'\Theta_1(\lambda',\gamma) -\lambda\Theta_1(\lambda,\gamma))}{(\lambda'-\lambda)}.\]
\end{proposition}

The proofs of these propositions are given in Section S.3.
Since $\bar{R}$ has finite fourth moment, it follows immediately from Propositions \ref{proposition:k_0} and \ref{proposition:asym_mean_variance} 
that
\[
d_{TV}(\bar{R},U)\leq\frac{2\sqrt{5}\{c_1c_2\varrho_0(\bar{R})+c_1^3\varrho_1(\bar{R})\varrho_2(\bar{R})\}}{\var(\bar{R})}\to 0,
\]
where $U$ is a normal random variable with the same mean and variance as $\bar{R}$.
The asymptotic normality of $\bar{R}$ now follows. From this, the asymptotic mean and variance of $\RHT(\lambda)$
follow from basic calculus, making use of the $\delta$-method and the relation shown in (\ref{eq:RHT_U}). Details are omitted. Finally we are able to conclude
\[\sqrt{p}\frac{\{p^{-1}\RHT(\lambda) -\Theta_1(\lambda,\gamma)\} }{\{2\Theta_2(\lambda,\gamma)\}^{1/2}}\Longrightarrow \mathcal{N}(0,1). \]

\subsection{Proof of Theorem \ref{thm:local_power}}

Under the deterministic local alternative, we denote $Y_{ij}=X_{ij}-\mu_i$. Then 
\[S_n=\frac{1}{n-2}\sum_{i=1}^2\sum_{j=1}^{n_i}Y_{ij}Y_{ij}^T-\frac{n_1}{n-2}\bar{Y}_1\bar{Y}_1^T-\frac{n_2}{n-2}\bar{Y}_2\bar{Y}_2^T.\]
Furthermore, redefine 
\[\tS=\frac{1}{n}\sum_{i=1}^2\sum_{j=1}^{n_i}Y_{ij}Y_{ij}^T.\]
With $g_n = {\kappa_{n}(1-\kappa_n)}\{2\gamma_n\hat{\Theta}_2(\lambda,\gamma_n)\}^{-1/2}$, the statistic under the local alternative can be written as 
\begin{align}
T_{n,p}(\lambda)=& T^{0}_{n,p}(\lambda) + g_n n^{1/2}\mu^T (S_n+\lambda I_p)^{-1}\mu \label{eq:exp_local_alternative}\\
& - 2g_n n^{1/2}\mu^T (S_n+\lambda I_p)^{-1} \bar{Y}_1  + 2g_n n^{1/2}\mu^T (S_n+ \lambda I_p)^{-1} \bar{Y}_2.\nonumber
\end{align}
where $T_{n,p}^0(\lambda)$ is the standardized statistic with $\{Y_{ij}\}$ as observations. We already proved $T_{n,p}^0(\lambda)$ converges to $\mathcal{N}(0,1)$ in distribution. 
To this end, it is enough to show that, under the stability condition (\ref{eq:mu_stability_condition}), 
\begin{align*}
&n^{1/2}\mu^T (S_n+\lambda I_p)^{-1} \mu -  q(\lambda,\gamma)  = o_p(1), \\
&n^{1/2}\mu^T (S_n+\lambda I_p)^{-1} \bar{Y}_i = o_p(1), \qquad i=1,2.
\end{align*}
Using the relation shown in (\ref{eq:S_ts}), we can write 
\begin{align*}
&n^{1/2} \mu^T (S_n+\lambda I_p)^{-1} \mu 
= n^{1/2}\mu^T(\tS+\lambda I_p)^{-1}\mu + 
n^{1/2}\Big(\begin{matrix} U_{\mu,1}, U_{\mu,2}\end{matrix}\Big) \mathbb{H}^{-1} \Big(\begin{matrix} 
U_{\mu,1} \\
U_{\mu,2}
\end{matrix} \Big),\\
&n^{1/2} \mu^T (S_n+\lambda I_p)^{-1} \bar{Y}_1
= n^{1/2} U_{\mu,1} + 
n^{1/2}\Big(\begin{matrix} U_{\mu,1}, U_{\mu,2}\end{matrix}\Big) \mathbb{H}^{-1} \Big(\begin{matrix} 
U_{11} \\
U_{12}
\end{matrix} \Big),\\
&n^{1/2} \mu^T (S_n+\lambda I_p)^{-1} \bar{Y}_2
= n^{1/2} U_{\mu,2} + 
n^{1/2}\Big(\begin{matrix} U_{\mu,1}, U_{\mu,2}\end{matrix}\Big) \mathbb{H}^{-1} \Big(\begin{matrix} 
U_{12} \\
U_{22}
\end{matrix} \Big),
\end{align*}
where $U_{ii'}$ and $\mathbb{H}$ are defined in the same way as in (\ref{eq:U_i_iprime_def}) and (\ref{eq:S_ts}), but with $X_{ij}$ replaced by $Y_{ij}$,  and $U_{\mu,i} = \mu^T(\tS+\lambda I_p)^{-1}\bar{Y}_i$, $i=1,2$. 

Proposition \ref{proposition:asym_mean_variance} implies that $U_{11}, U_{12}, U_{22}$ converge in probability to deterministic quantities and $\mathbb{H}$ converges in probability to a nonsingular matrix.  
Therefore, it suffices to show 
\begin{align}
&n^{1/2}\mu^T (\tS+\lambda I_p)^{-1} \mu -  q(\lambda,\gamma)  = o_p(1),\label{eq:local_power_term1} \\
&n^{1/2}\mu^T (\tS+\lambda I_p)^{-1} \bar{Y}_i = o_p(1),\quad i =1,2.\label{eq:local_power_term2}
\end{align}
Equation (\ref{eq:local_power_term1}) is a special case of the limiting behavior of quadratic forms considered by \cite{ElKaroui2011}, and its proof follows along the material in Section 2 and Section 3 of their paper. The proof of (\ref{eq:local_power_term2}) is given in Section S.3.5 of the Supplementary Material. 

\subsection{Proof of Theorem \ref{thm:local_power_probabilistic}}
Under the prior distribution given by \textbf{PA}, decompose $T_{n,p}(\lambda)$ as
\begin{equation*}
T_{n,p}(\lambda) = T_{n,p}^0(\lambda) + g q(\lambda,\gamma) +\sigma_n(\mu)+ \sum_{i=1}^2\eta_n^{(i)}(Y) + \sum_{j=1}^4 \delta_n^{(j)}(\mu, Y),
\end{equation*}
where, with $g = \kappa(1-\kappa)\{2\gamma\Theta_2(\lambda,\gamma)\}^{-1/2}$,
\begin{align*}
\sigma_n(\mu)&= g[n^{1/2}\mu^T D(-\lambda)\mu -p^{-1}\tr(D(-\lambda)\mathbf{B})],\\[5pt]
\eta_n^{(1)}(Y) &= (g_n-g)q(\lambda,\gamma), \\[5pt]
\eta_n^{(2)}(Y) &= g_n[p^{-1}\tr(D(-\lambda)\mathbf{B})- q(\lambda,\gamma)],\\[5pt]
\delta_n^{(1)}(\mu,Y)&= (g_n-g)[n^{1/2}\mu^T D(-\lambda)\mu -p^{-1}\tr(D(-\lambda)\mathbf{B})],\\[5pt]
\delta_n^{(2)}(\mu,Y)&= g_n[n^{1/2}\mu^T(S_n+\lambda I_p)^{-1}\mu - n^{1/2}\mu^T D(-\lambda)\mu],\\[5pt]
\delta_n^{(3)}(\mu,Y)&= g_n n^{1/2}\mu^T(S_n+\lambda I_p)^{-1}\bar{Y}_1,\\[5pt]
\delta_n^{(4)}(\mu,Y)&= g_n n^{1/2}\mu^T(S_n+\lambda I_p)^{-1}\bar{Y}_2.
\end{align*}
Through this subsection, we use $\mathbb{P}_*$ to mean the prior probability measure of $\mu$ and use $\mathbb{P}_{\mu}$ to mean the probability of $X_{ij}$ conditional on $\mu$.  
The power under the alternative $\mu$ is then
\[\beta_n(\mu,\lambda) = \mathbb{P}_{\mu}\{ T_{n,p}(\lambda) >\xi_{\alpha} \}. \]

To show (\ref{eq:RHT_asymptotic_power_probabilistic}),  it suffices to show that for any $\epsilon>0$ and any $\zeta>0$, there exists a sufficiently large $N$, such that when $n>N$,
\[\mathbb{P}_*\Big(\Big| \beta_n(\mu,\lambda) -\Phi(-\xi_\alpha + g q(\lambda,\gamma)) \Big|>\epsilon\Big)<\zeta.\]

Due to Lemma \ref{lemma:Bai&Silverstein} and the assumption $\mu = n^{-1/4} p^{-1/2} B\nu$, 
\[n^{1/2} \mu^T D(-\lambda)\mu - p^{-1} \tr(D(-\lambda)BB^T) \stackrel{\mathbb{P}_*}{\longrightarrow} 0.  \]
Therefore, there exist a constant $C_\epsilon$ and a sufficiently large $N_1$ such that when $n>N_1$,
\[\mathbb{P}_*(K^{(1)}_\epsilon) \geq 1- \zeta,\]
where 
\[K_\epsilon^{(1)} = \{\mu: n^{1/2} \|\mu\|^2 \leq C_\epsilon\}\cap \{\mu: |\sigma_n(\mu)|\leq \epsilon\}.\]
Next, $g_n$ is independent with $\mu$ and as introduced in Section \ref{sec:RHT:prelim},
\[g_n \stackrel{\mathbb{P}_{\mu}}{\longrightarrow} g, \qquad \mbox{as } n,p\to \infty. \]
Therefore, when $\mu \in K_\epsilon^{(1)}$, as $n,p\to\infty$, with a tail bound not depending on $\mu$,
\begin{align*}
\max\limits_{i=1,2}|\eta_n^{(i)}(Y)| \stackrel{\mathbb{P}_{\mu}}{\longrightarrow} 0,\qquad \mbox{and}\qquad |\delta_n^{(1)}(\mu,Y)| \stackrel{\mathbb{P}_{\mu}}{\longrightarrow} 0. 
\end{align*}
As for $\delta_n^{(j)}(\mu,Y), j=2,3,4$, arguments analogous to those in Theorem 3.1 and Proposition 3.1 of \cite{ElKaroui2011} show that, as $n,p\to\infty$,
\[n^{1/2}\mu^T(\tS +\lambda I_p)^{-1}\mu - n^{1/2}\mu^T D(-\lambda)\mu  \stackrel{\mathbb{P}_\mu}{\longrightarrow} 0,  \] 
with a tail bound only depending on $n^{1/2}\|\mu\|^2$.  Moreover, the proof of Theorem \ref{thm:local_power_probabilistic} shows
\begin{align*}
n^{1/2} \mu^T (\tS+\lambda I_p)^{-1} \bar{Y}_i \stackrel{\mathbb{P}_\mu}{\longrightarrow}0, \quad i=1,2,   
\end{align*} 
also with a tail bound only depending on $n^{1/2}\|\mu\|^2$ (see Section S.3.5 of the Supplementary Material). Together with the relation shown in (\ref{eq:S_ts}), we conclude that on $\mu \in K_\epsilon^{(1)}$, with an uniform tail bound, as $n,p\to\infty$,
\[ \max\limits_{j=2,3,4} |\delta_n^{(j)}(\mu,Y)| \stackrel{\mathbb{P}_\mu}{\longrightarrow}0.\]
The analysis up to now implies that we can find a sufficiently large $N_2$ such that when $n>N_2$, 
\begin{align*}
&\mathbb{P}_{\mu}( K_\epsilon^{(2)} )> 1-\epsilon,
\end{align*}
for any $\mu\in K_\epsilon^{(1)}$, where  
\[K_\epsilon^{(2)} = K_\epsilon^{(1)} \cap \{Y_{ij}\colon \max\limits_{i=1,2} |\eta_n^{(i)}(Y)| \leq \epsilon\quad \mbox{and}\quad \max\limits_{i=1,2,3,4} |\delta_n^{(i)}(\mu,Y)| \leq \epsilon \}.\]
Since
\[\mathbb{P}_\mu(T_{n,p}(\lambda)>\xi_\alpha) =  \mathbb{P}_\mu(\{T_{n,p}(\lambda)>\xi_\alpha\} \cap  K_\epsilon^{(2)} )+ \mathbb{P}_\mu(\{T_{n,p}(\lambda)>\xi_\alpha\} \cap \{ K_\epsilon^{(2)}\}^c ),  \]
it follows that
\begin{align*}
&\mathbb{P}_\mu(T_{n,p}(\lambda)>\xi_\alpha) \leq \epsilon + \mathbb{P}_\mu(T_{n,p}^0(\lambda) > \xi_\alpha- g q(\lambda,\gamma)-7\epsilon),\\
&\mathbb{P}_\mu(T_{n,p}(\lambda)>\xi_\alpha) \geq -\epsilon + \mathbb{P}_\mu(T_{n,p}^0(\lambda) > \xi_\alpha -g q(\lambda,\gamma) +7\epsilon).
\end{align*}
On the other hand, since $T_{n,p}^0(\lambda)$ is free of $\mu$ and converges in distribution to standard normal distribution, we can find a sufficiently large $N_3$ such that when $n>N_3$, for any $\mu\in K^{(1)}_\epsilon$,
\begin{align*}
P_{\mu} (T_{n,p}^0(\lambda) > \xi_\alpha - g q(\lambda,\gamma) - 7\epsilon) &< \Phi(-\xi_\alpha +gq(\lambda,\gamma)-7\epsilon)+\epsilon\\
P_{\mu} (T_{n,p}^0(\lambda) > \xi_\alpha - g q(\lambda,\gamma) + 7\epsilon) &> \Phi(-\xi_\alpha +gq(\lambda,\gamma)+ 7\epsilon)-\epsilon.
\end{align*}
In summary, on $\mu\in K_\epsilon^{(1)}$, when $n>\max\limits_{i=1,2,3} N_{i}$,
\begin{align*}
\mathbb{P}_\mu(T_{n,p}(\lambda)>\xi_\alpha) &\leq 2\epsilon + \Phi(-\xi_\alpha +gq(\lambda,\gamma)-7\epsilon),\\
\mathbb{P}_\mu(T_{n,p}(\lambda)>\xi_\alpha) &\geq -2\epsilon + \Phi(-\xi_\alpha +gq(\lambda,\gamma)+7\epsilon).
\end{align*}
This completes the proof, since $\mathbb{P}_*(K_\epsilon^{(1)}) \geq 1-\zeta.$

\newpage
\subsection{Proof of Theorem \ref{thm:converg_lambda}}
\label{subsec:proof_theorem_converg_lambda}

\subsubsection{Proof of ~(\ref{eq:converg_lambda})}
To show the existence of a sequence of local maximizers of $\hat{Q}_n(\lambda,\gamma_n)$ as stated,
it suffices to show that for any $\varepsilon\in(0,1)$, there exists a constant $K>0$, and an
integer $n_\varepsilon$, such that, for $t=Kn^{-1/4}$,
\[\mP\left\{\hat{Q}_n(\lambda_\infty\pm t,\gamma_n)-\hat{Q}_n(\lambda_\infty,\gamma_n)\leq0\right\}\geq\varepsilon\]
for all $n\geq n_\varepsilon$. If we use a stochastic term $\delta(t)$ to measure the difference between $\hat{Q}_n(\lambda,\gamma_n)$ and $Q(\lambda,\gamma)$ at $\lambda=\lambda_\infty\pm t$ and $\lambda_\infty$, considering $\lambda_\infty$ to be in the interior of $[\underline{\lambda},\overline{\lambda}]$, a second-order Taylor expansion yields
\begin{align*}
\hat{Q}_n(\lambda_\infty\pm t,\gamma_n)-\hat{Q}_n(\lambda_\infty,\gamma_n) &=Q(\lambda_\infty\pm t,\gamma)-Q(\lambda_\infty,\gamma)+\delta(\pm t)\\
&=\frac{t^2}{2}\frac{\partial^2}{\partial\lambda^2}Q(\lambda_\infty,\gamma)+O(t^3)+\delta(\pm t)
\end{align*}
Since $O(t^3)$ is a smaller order term as $n\to\infty$ and
${\partial^2}Q(\lambda_\infty,\gamma)/{\partial\lambda^2}<0$, it suffices to
show that ${n^{1/2}}|\delta(\pm t)|=O_p(1)$ with an uniform tail bound in $t$. Again by Taylor expansion,
\begin{align*}
{n^{1/2}}\delta(\pm t)=&{n^{1/2}}t\Big[\frac{\partial}{\partial\lambda}\hat{Q}_n(\lambda_\infty,\gamma_n)
-\frac{\partial}{\partial\lambda}Q(\lambda_\infty,\gamma)\Big]\\
&\quad 
+\frac{{n^{1/2}}t^2}{2}\Big[\frac{\partial^2}{\partial\lambda^2}\hat{Q}_n(\lambda_\infty,\gamma_n)
-\frac{\partial^2}{\partial\lambda^2}Q(\lambda_\infty,\gamma)\Big]\\
&\quad +\frac{{n^{1/2}}t^3}{6}\frac{\partial^3}{\partial\lambda^3}\hat{Q}(\lambda_\infty+\alpha t,\gamma_n)-\frac{{n^{1/2}}t^3}{6}\frac{\partial^3}{\partial\lambda^3}Q(\lambda_\infty+\alpha t,\gamma)
\end{align*}
for some $\alpha\in[0,1]$. 

Now expressing $Q(\lambda,\gamma)$ $\hat Q_n(\lambda,\gamma)$ and 
their partial derivatives as continuous functions of $m_F(-\lambda_\infty)$, $m_F^\prime(-\lambda_\infty)$, $m_F^{(3)}(-\lambda_\infty)$, $m_F^{(4)}(-\lambda_\infty)$,$\phi$,$\gamma$,
and their empirical counterparts, we use Proposition \ref{prop:converg_phi}--\ref{prop:coverg_mF} to deduce that
\begin{align*}
n^{1/4}&\Big|\frac{\partial}{\partial\lambda}\hat{Q}_n(\lambda_\infty,\gamma)
-\frac{\partial}{\partial\lambda}Q(\lambda_\infty,\gamma)\Big|\stackrel{P}{\longrightarrow}0,\\
&\Big|\frac{\partial^2}{\partial\lambda^2}\hat{Q}_n(\lambda_\infty,\gamma)
-\frac{\partial^2}{\partial\lambda^2}Q(\lambda_\infty,\gamma)\Big|\stackrel{P}{\longrightarrow}0,\\
\sup\limits_{\lambda\in[\underline{\lambda},\overline{\lambda}]}
&\Big|\frac{\partial^3}{\partial\lambda^3}\hat{Q}_n(\lambda,\gamma)\Big|
+\Big|\frac{\partial^3}{\partial\lambda^3}Q(\lambda,\gamma)\Big|=O_p(1).
\end{align*}
which completes the proof. If $\lambda_\infty$ is on the boundary and
$\partial Q(\lambda_\infty,\gamma)/{\partial\lambda}<0$, similar results follow from a first-order Taylor expansion.

\subsubsection{Proof of ~(\ref{eq:normality_lambda_n})}
It remains to verify (\ref{eq:normality_lambda_n}). To this end, note that it suffices to prove that
\begin{align*}
&{p^{1/2}}\Big|\frac{1}{p}\RHT(\lambda_n)-\hat{\Theta}_1(\lambda_n,\gamma_n)
-\frac{1}{p}\RHT(\lambda_\infty)+\hat{\Theta}_1(\lambda_\infty,\gamma_n)\Big|\\
&\leq{p^{1/2}}\Big|\frac{1}{p}\frac{\partial}{\partial\lambda}\RHT(\lambda_\infty)
-\frac{\partial}{\partial\lambda}\hat{\Theta}_1(\lambda_\infty,\gamma_n)\Big||\lambda_n-\lambda_\infty|\\
&+\frac{{p^{1/2}}}{2}\Big|\frac{1}{p}\frac{\partial^2}{\partial\lambda^2}\RHT(\lambda_\infty)
-\frac{\partial^2}{\partial\lambda^2}\hat{\Theta}_1(\lambda_\infty,\gamma_n)\Big||\lambda_n-\lambda_\infty|^2\\
&+\frac{{p^{1/2}}}{6}\Big|\frac{1}{p}\frac{\partial^3}{\partial\lambda^3}\RHT(\lambda^*)
-\frac{\partial^3}{\partial{\lambda^3}}\hat{\Theta}_1(\lambda^*,\gamma_n)\Big|
|\lambda_n-\lambda_\infty|^3\stackrel{P}\longrightarrow0
\end{align*}
where $\lambda^*$ is in between $\lambda_\infty$ and $\lambda_n$.  So it is enough to show that
\begin{align}
p^{1/4}&\Big|\frac{1}{p}\frac{\partial}{\partial\lambda}\RHT(\lambda_\infty)
-\frac{\partial}{\partial\lambda}\hat{\Theta}_1(\lambda_\infty,\gamma_n)\Big|
\stackrel{P}\longrightarrow0,\label{eq:first_order_converg}\\
&\Big|\frac{1}{p}\frac{\partial^2}{\partial\lambda^2}\RHT(\lambda_\infty)
-\frac{\partial^2}{\partial\lambda^2}\hat{\Theta}_1(\lambda_\infty,\gamma_n)\Big|
\stackrel{P}\longrightarrow0,\label{eq:second_order_converg}\\
\sup\limits_{\lambda\in[\underline{\lambda},\overline{\lambda}]}
&\Big|\frac{1}{p}\frac{\partial^3}{\partial\lambda^3}\RHT(\lambda)
-\frac{\partial^3}{\partial{\lambda^3}}\hat{\Theta}_1(\lambda,\gamma_n)\Big|
=O_p(1).\label{eq:third_order_converg}
\end{align}
Next, 
\[
\mE\left|p^{-1}\frac{\partial^3}{\partial\lambda^3}\RHT(\lambda)\right|
	\leq\frac{n_1n_2}{\underline{\lambda}^{-4}p(n_1+n_2)}\mE|(\bar{X}_1-\bar{X}_2)^T(\bar{X}_1-\bar{X}_2)|
	=O(1)
\]
for all
$\lambda\in[\underline{\lambda},\overline{\lambda}]$. And Proposition \ref{prop:coverg_mF} shows the convergence of ${\partial^3}\hat{\Theta}_1(\lambda,\gamma_n)/{\partial\lambda^3}$ to
${\partial^3\Theta_1(\lambda,\gamma)}/{\partial\lambda^3}$ uniformly on $\lambda\in[\underline{\lambda},\overline{\lambda}]$,
so that (\ref{eq:third_order_converg}) holds.
	
For proving (\ref{eq:first_order_converg}) and (\ref{eq:second_order_converg}), note that Propositions \ref{prop:first_deriv_Theta1} and \ref{prop:second_deriv_Theta1} showed the convergence of ${\partial}\hat{\Theta}_1(\lambda,\gamma_n)/{\partial\lambda}$ to $-{p^{-1}}\tr\left[\{R_n(-\lambda)\}^2\Sigma_p\right]$,  and the convergence of ${\partial^2}\hat{\Theta}_1(\lambda,\gamma_n)/{\partial\lambda^2}$ to
${2}{p^{-1}}\tr\left[\{R_n(-\lambda)\}^3\Sigma_p\right]$.
So the proof will be complete if we can show
\begin{align}
p^{1/4}&\Big|\frac{1}{p}\frac{\partial}{\partial\lambda}\RHT(\lambda_\infty)
+\frac{1}{p}\tr\left[\{R_n(-\lambda_\infty)\}^2\Sigma_p\right]\Big|\stackrel{P}\longrightarrow 0,
\label{eq:first_deriv_RHT}\\
&\Big|\frac{1}{p}\frac{\partial^2}{\partial\lambda^3}\RHT(\lambda_\infty)-\frac{2}{p}
\tr\left[\{R_n(-\lambda_\infty)\}^3\Sigma_p\right]\Big|
\stackrel{P}\longrightarrow0.\label{eq:second_deriv_RHT}
\end{align}
We move the proofs of (\ref{eq:first_deriv_RHT}) and (\ref{eq:second_deriv_RHT}) to Section S.3.6 and S.3.7 of the Supplementary Material, which are lengthy.

\subsection{Proof of Theorem \ref{thm:convergence_process}}
\label{sec:derivation_gamma}
To prove the process convergence stated in Theorem \ref{thm:convergence_process}, we need to verify the convergence
of finite-dimensional distributions and the tightness of the 
process.

(a) To show the distributional convergence of $\{\RHT(\lambda_1),\dots,\RHT(\lambda_k)\}$ for arbitrary integer $k$
and fixed $\lambda_1,\ldots,\lambda_k > 0$, it suffices to show the joint normality of $\{U_{ii'}(\lambda_j), 1\leq i,i'\leq 2, 1\leq j\leq k\}$. Therefore, define an arbitrary linear combination 
\[
T_n=\sum_{i=1}^2\sum_{i'=1}^2\sum_{j=1}^k l_{ii'j}U_{ii'}(\lambda_j)
\]
It suffices to show that $T_n$ is asymptotically normal. We can derive asymptotic orders of the functions $\varrho_0$, $\varrho_1$ and $\varrho_2$ with each $U_{ii'}(\lambda_j)$ as arguments and combine them through Cauchy-Schwarz inequality to get the asymptotic orders of $\varrho_0$, $\varrho_1$, $\varrho_2$ with $T_n$ as the argument. The proof is essentially
a repetition of the arguments in Section \ref{subsec:asy_normality},
and is hence omitted.

(b) To show tightness, note first that Proposition \ref{prop:coverg_mF} yields
$\hat\Theta_2(\lambda,\gamma_n)\to_p\Theta_2(\lambda,\gamma)$ uniformly on $[\underline{\lambda},\bar{\lambda}]$.
This implies tightness of $(\hat{\Theta}_2(\lambda,\gamma_n)\colon\lambda\in[\underline{\lambda},\bar{\lambda}])$.
The sequence ${n^{1/2}}({p^{-1}}\RHT(\lambda)-\hat{\Theta}_1(\lambda,\gamma_n))$ is shown to be
tight in \citet[Section 4]{PanZhou2011} for observations with finite fourth moments but with $\Sigma = I_p$.
Although their arguments are in a one-sample testing framework, they can easily be generalized to
the two-sample testing case and for $\Sigma$ satisfying \textbf{C1}--\textbf{C3}. Together with $\inf_{\lambda\in[\underline{\lambda},\bar{\lambda}]}\Theta_2(\lambda,\gamma)>0$,
the convergence of the process follows. 

(c) The covariance kernel can be computed via basic calculus, making use of Proposition \ref{proposition:asym_mean_variance} and the relation between $\bar{R}(\lambda)$ and $\RHT(\lambda)$ shown in (\ref{eq:RHT_U}).

\subsection{Proof of Proposition \ref{prop:RHT_LAM}}\label{subsec:proof_RHT_LAM}

{In order to find the minimax rule within
	$\mathcal{D}$, we first find $\tilde{\pi}_\lambda$ 
	which minimizes $Q(\lambda,\gamma;\tilde{\pi})$ for $\tilde{\pi} \in \Pi_2(1)$, for every fixed $\lambda$. At this point we make two important
	observations:
	\begin{itemize}
		\item[(i)] $\Pi_2(1)$ is convex.
		\item[(ii)] $(0,0,1/\phi_2)$ is an extreme point of $\Pi_2(1)$, while 
		$\pi_0 \geq 0$ and $\pi_2 \geq 0$ for all $\tilde{\pi} = (\pi_0,\pi_1,\pi_2) \in \Pi_2(1)$.
	\end{itemize}
	Because of (i), and the fact that $Q(\lambda,\gamma;\tilde{\pi})$ is linear in $\tilde{\pi}$, the minimum occurs at the boundary of the 
	set $\Pi_2(1)$.

	The following proposition establishes that $\tilde\pi = \phi_2^{-1}\mathbf{e}_2$, where $\mathbf{e}_2 = (0,0,1)$.
	\begin{proposition}\label{prop:rho_inequality}
		For $j=0,1,\ldots$, 
		\begin{equation}\label{eq:rho_inequality}
		\phi_j^{-1}\rho_j(-\lambda,\gamma) \geq \phi_{j+1}^{-1}\rho_{j+1}(-\lambda,\gamma), \qquad 
		\mbox{for all}~\lambda > 0,
		\end{equation}
		where $\phi_j = \int \tau^j dH(\tau)$.
	\end{proposition}	
	To verify the claim that $\tilde\pi = \phi_2^{-1}\mathbf{e}_2$, observe that minimization of $Q(\lambda,\gamma;\tilde{\pi})$ is equivalent to minimization of 
	$\sum_{j=0}^2 \pi_j \rho_j(-\lambda,\gamma)$
	over $\tilde{\pi}\in \Pi_2(1)$. Using the fact that $\phi_0 = 1$, for any $\tilde{\pi} \in \Pi_2(1)$,
	\begin{align*}
	 	\sum_{j=0}^2 \pi_j& \rho_j(-\lambda,\gamma) - \phi_2^{-1} \rho_2(-\lambda,\gamma) \\
		=& \pi_0 (\phi_0^{-1}\rho_0(-\lambda,\gamma) -  
		\phi_1^{-1}\rho_1(-\lambda,\gamma)) \\
		&+ (1-\phi_2\pi_2) (\phi_1^{-1}\rho_1(-\lambda,\gamma)-\phi_2^{-1}\rho_2(-\lambda,\gamma)),
	\end{align*}
	which follows from substituting $\phi_1 \pi_1 = 1 - \pi_0 - \phi_2\pi_2$. 
	Now by (ii) and Proposition \ref{prop:rho_inequality}, the right hand side is nonnegative, and equals zero only if $\tilde{\pi} = \phi_2^{-1}\mathbf{e}_2$, which verifies the claim.

The next step is therefore to find $\lambda \in [\underline{\lambda},\overline{\lambda}]$ that maximizes 
	$Q(\lambda,\gamma;\phi_2^{-1}\mathbf{e}_2) = \phi_2^{-1} Q(\lambda,\gamma;\mathbf{e}_2)$. Due to Proposition \ref{prop:Q2_monotonic}, stated below, the maximum occurs at $\lambda = 
	\overline{\lambda}$, which shows that $T_{n,p}(\overline{\lambda})$ is 
	LAM with respect the class $\mathfrak{P}_{2}(C)$ for any $C > 0$.
	\begin{proposition}\label{prop:Q2_monotonic}
		The function $Q(\lambda,\gamma;\mathbf{e}_2)$ is nondecreasing
		on $[\underline{\lambda},\infty)$ for any $\underline{\lambda} > 0$, where
		$\mathbf{e}_2 = (0,0,1)$. 
	\end{proposition}
	Proof of Propositions \ref{prop:rho_inequality} and \ref{prop:Q2_monotonic}
	are given in the Supplementary Material.


\setcounter{section}{0}
\renewcommand{\thesection}{A.\arabic{section}}
\setcounter{equation}{0}
\renewcommand{\theequation}{A.\arabic{equation}}
\renewcommand{\thesubsection}{A.\arabic{section}.\arabic{subsection}}
\setcounter{table}{0}
\renewcommand{\thetable}{A.\arabic{table}}
\setcounter{figure}{0}
\renewcommand{\thefigure}{A.\arabic{figure}}

\setcounter{theorem}{0}
\renewcommand\thetheorem{A.\arabic{theorem}}
\setcounter{lemma}{0}
\renewcommand\thelemma{A.\arabic{lemma}}
\setcounter{proposition}{0}
\renewcommand\theproposition{A.\arabic{proposition}}
\renewcommand\theclaim{S.\arabic{section}.\arabic{claim}}

\section*{Appendix}

\subsection*{Technical tools}
\label{subsec:technical_lemmas}
There are a collection of lemmas and propositions whose proofs are gathered in Section S.3 of the Supplementary Material.  In what follows, let $\|\cdot\|$
be the operator norm of a matrix and $\|\cdot\|_F$ denote the Frobenius norm.

\begin{lemma}
	\label{lemma: She_Mor}
	(Sherman--Morrison Formula).
	Suppose $A$ is an invertible square matrix and $u,v$ are column vectors. Suppose furthermore that $1+v^TA^{-1}u\neq0$. Then
	\begin{equation}
	\label{eq:She_Mor}
	(A+uv^T)^{-1}=A^{-1}-\frac{A^{-1}uv^TA^{-1}}{1+v^TA^{-1}u}.
	\end{equation}
\end{lemma}

\begin{lemma}
	\label{lemma:bounded_diff}
	Suppose we have two matrices $A$ and $B$ with $A$ symmetric and positive definite. For any vector $Y$ and any integer $k\geq1$, 
	\begin{equation*}
	\left|\tr\{(A+YY^T)^{-k}B\}-\tr(A^{-k}B)\right|\leq \frac{k\|B\|}{\tau_A^k},
	\end{equation*}  
	where $\tau_A$ is the smallest eigenvalue of $A$.
\end{lemma}

\begin{lemma}
	\label{lemma:Hanson-Wright}
	(Hanson--Wright inequality). Let\/ $Y=(Y_1,\ldots,Y_n)^T\in \mR^n$ be a random vector
	with independent components $Y_i$ having $\mE[Y_i]=0$ and uniformly bounded $\psi_2$-norm (sub-Gaussian norm)
	\[
	\|Y_i\|_{\psi_2}
	=\sup\limits_{p\geq1}\frac{1}{p^{1/2}}\big(\mE\big[|Y_i|^p\big]\big)^{1/p}
	\leq K,
	\qquad i=1,\ldots,n,
	\]
	where $K>0$ is a constant. Let $A$ be an $n\cdot n$ matrix. Then, for any $t\geq0$,
	\[
	\mP\left(\big|Y^TAY-\mE[Y^TAY]\big|>t\right)
	\leq2\exp\left\lbrace-c\min\left(\frac{t^2}{K^4\|A\|_{F}^2},\frac{t}{K^2\|A\|}\right)\right\rbrace,
	\]
	where $c>0$ is a constant.
\end{lemma}

\begin{lemma}
	\label{thm:nonasy_eigen}
	(Theorem 5.39 of \cite{Vershynin2012}).
	Let $\X$ be an $n \times p$ matrix whose rows $X_i$ are independent sub-gaussian isotropic random vectors in $\mR^n$. Then for every $t\geq0$, with probability at least $1-2\exp(-ct^2)$ one has
	\begin{equation*}
	{n}^{1/2}-C{p}^{1/2}-t\leq s_{\min}(\X)\leq s_{\max}(\X)\leq{n}^{1/2}+C{p}^{1/2}+t
	\end{equation*}
	where $s_{\min}$ and $s_{\max}$ are the smallest and largest singular value of $\X$, and $C=C_K$, $c=c_K>0$ depend only on the subgaussian norm $K=\max_i\|X_i\|_{\psi_2}$ of the rows.
\end{lemma}

\begin{lemma}
	\label{lemma:Bai&Silverstein}(Lemma 2.7 of~\cite{BaiSilverstein1998}).
	Let $Z=(z_1,\dots,z_p)^T$, where $z_i$'s are independent random variables with mean 0 and variance 1. Let $A$ be a deterministic  matrix. Then for any $k\geq2$, we have
	\begin{equation}
	\label{eq:Bai&Silverstein}
	\mE|Z^TAZ-\tr A|^k\leq C_k\left[\{\mE z_1^4\tr(AA^T)\}^{k/2}+\mE z_1^{2k}\tr\{(AA^T)^{k/2}\}\right].
	\end{equation}
	where $C_k$ is a constant depending on k only.
\end{lemma}

\begin{lemma}
	\label{lem:spectral_moment}
	(Lemma 1 of ~\cite{BaiCY2010}).
	Let $F_{\gamma,H}$ denote the limiting empirical spectral distribution of 
	$S_n$. Then, under conditions \textbf{C1}--\textbf{C3}, the moments
	$\theta_j = \int x^j dF_{\gamma,H}(x)$ of 
	$F_{\gamma,H}$ are linked to the moments $\phi_j = \int \tau^j dH(\tau)$ of the population spectral distribution $H$ by
	\begin{equation*}
	\theta_j = \sum \gamma^{(i_1+i_2+ \cdots+ i_j)-1} (\phi_1)^{i_1}
	(\phi_2)^{i_2}\cdots(\phi_j)^{i_j} K_{i_1,i_2,\ldots,i_j}^{(j)},
	\end{equation*}
	where the sum runs over the following partition of $j$:
	\begin{equation*}
	(i_1,\ldots,i_j):~j = i_1 + 2i_2 + \cdots + j i_j, \qquad i_\ell \in \mathbb{N},
	\end{equation*}
	and 
	\begin{equation*}
	K_{i_1,i_2,\ldots,i_j}^{(j)} = \frac{j!}{i_1!i_2!\cdots i_j!(j-i_1-\cdots-i_j)!}~.
	\end{equation*}
\end{lemma}

\begin{theorem}
	\label{thm:Chatterjee2009}
	(Theorem 2.2 of~\cite{Chatterjee2009}).
	Let $\Z=(z_1,\dots,z_n)$ be a vector of independent random variables in $\mathcal{L}(c_1,c_2)$ for some finite $c_1,c_2$. Take any $g\in C^2(\mR^n)$ and let $\nabla g$ and $\nabla^2 g$ denote the gradient and Hessian of $g$. Let
	\begin{align*}
	&\varrho_0(g)=\left(\mE\sum\limits_{i=1}^n\left|\frac{\partial g}{\partial z_i}(\Z)\right|^4\right)^{1/2},\\
	&\varrho_1(g)=\left(\mE\|\nabla g(\Z)\|^4\right)^{1/4},\\
	&\varrho_2(g)=(\mE\|\nabla^2g(\Z)\|^4)^{1/4},
	\end{align*}
	where $\|\cdot\|$ is the operator norm.
	Suppose $W=g(\Z)$ has a finite fourth moment and let $\sigma^2=\var(W)$. Let $U$ be a normal random variable having the same mean and variance as $W$. Then
	\begin{equation}
	\label{eq:total_variation}
	d_{TV}(W,U)\leq\frac{2\sqrt{5}\{c_1c_2\varrho_0(g)+c_1^3\varrho_1(g)\varrho_2(g)\}}{\sigma^2},
	\end{equation}
	where $d_{TV}$ is the total variation distance between two distributions.
\end{theorem}

\subsection*{Key propositions used in the proofs}

In the following, $c_1$, $c_2$ and $c_3$ denote some universal positive constants, independent of
$\lambda$. To lighten notation, some fixed parameters are ignored in the following expressions
when it does not cause ambiguity; for example, weights $\widetilde{\pi}$ in $Q(\lambda,\gamma;\widetilde{\pi})$
may be dropped. The following propositions show the concentration 
of some quantities.
Recall that $\hat{\phi}_1=p^{-1}\tr(S_n)$ and $\phi_1=\int\tau dH(\tau)$.
\begin{proposition}
	\label{prop:converg_phi}
	If conditions~\textbf{C1}--\textbf{C3}
	are satisfied, then for any $t>0$,
	\begin{align*}
	&\mP\left\lbrace\left|\hat{\phi}_1-\mE\hat{\phi}_1\right|>t\right\rbrace\leq c_1\exp\{-\min(c_2nt^2,c_3nt)\}.
	\end{align*}
	Moreover, $\sqrt{n}\left|\mE\hat{\phi}_1-\phi_1\right|\to0$, as $n\to\infty$, since $\mE\hat{\phi}_1=\int\tau dH_p(\tau)$.
\end{proposition}

\begin{proposition}
	\label{prop:coverg_mF}
	Define $\mF^{(k)}(-\lambda)$ to be the $k$-th order derivative of $\mF(-\lambda)$ and $m_F^{(k)}(-\lambda)$
	to be the $k$-th order derivative of $m_F(-\lambda)$. If
	conditions \textbf{C1}--\textbf{C3} 
	are satisfied, then
	for any $t>0$, integer $k$ and $\lambda\in[\underline{\lambda},\overline{\lambda}]$,
	\[\mP\left\lbrace\left|\mF^{(k)}(-\lambda)-\mE\mF^{(k)}(-\lambda)\right|>t\right\rbrace\leq c_1\exp(-c_2nt^2).\]
	Moreover,
	\[{n^{1/2}}\left|\mE\mF^{(k)}(-\lambda)-m_F^{(k)}(-\lambda)\right|\to0.\]
	It follows, as continuous and monotone functions in $\lambda$,
	\[\sup\limits_{\lambda\in[\underline{\lambda},\overline{\lambda}]}
	\left|\mF^{(k)}(-\lambda)-m_F^{(k)}(-\lambda)\right|\stackrel{P}{\longrightarrow}0.\]
\end{proposition}

\begin{proposition}
	\label{prop:first_deriv_Theta1}
	If conditions \textbf{C1}--\textbf{C3}
	are satisfied, then for any $\lambda\in[\underline{\lambda},\overline{\lambda}]$,
	\[\frac{\partial}{\partial\lambda}\hat{\Theta}_1(\lambda,\gamma_n)
	= -\frac{1}{p}\tr\left[\{R_n(-\lambda)\}^2\Sigma_p\right]+o_p(n^{-1/4}).\]
\end{proposition}

\begin{proposition}
	\label{prop:second_deriv_Theta1}
	If conditions \textbf{C1}--\textbf{C3}
	are satisfied, then for any $\lambda\in[\underline{\lambda},\overline{\lambda}]$,
	\[\frac{\partial^2}{\partial\lambda^2}\hat{\Theta}_1(\lambda,\gamma_n)
	=\frac{2}{p}\tr\left[\left\{R_n(-\lambda)\right\}^3\Sigma_p\right]+o_p(1).\]
\end{proposition}

\begin{proposition}
\label{prop:covariance_expression}
If conditions \textbf{C1}--\textbf{C3} are satisfied, then for any $\lambda,\lambda'\in[\underline{\lambda}, \overline{\lambda}], \lambda\neq \lambda'$,
\begin{align*}&
\frac{1}{p} \tr[R_n(-\lambda)\Sigma_p R_n(-\lambda')\Sigma_p]\\
&= \{1+\gamma\Theta_1(\lambda,\gamma)\}
\{1+\gamma\Theta_1(\lambda^\prime,\gamma)\}\left\{\frac{\lambda^\prime\Theta_1(\lambda^\prime,\gamma)
	-\lambda\Theta_1(\lambda,\gamma)}{\lambda^\prime-\lambda}\right\} +o_p(1).
\end{align*}
\end{proposition}


\section*{Supplementary material}
\label{SM}
Supplementary Material includes additional simulation results and detailed
proofs of the main theoretical results presented in this paper.

\newpage

\end{document}